\documentclass[lettersize,journal]{IEEEtran}

\usepackage{amsmath,amsfonts}

\usepackage{array}
\usepackage{textcomp}
\usepackage{stfloats}
\usepackage{url}
\usepackage{verbatim}
\usepackage{cite}

\usepackage{makecell}
\usepackage{float}
\usepackage{multicol}
\usepackage{amsmath,amssymb}
\usepackage{subfigure}
\usepackage{stfloats}
\usepackage{flushend}

\usepackage{balance}

\usepackage{amsthm}
\usepackage{makecell}
\usepackage{float}
\usepackage{graphicx,times,amsmath,booktabs}
\usepackage{multicol}

\usepackage{epstopdf}
\usepackage{stfloats}

\usepackage{balance}
\usepackage{xcolor}

\newtheorem{theorem}{Theorem}
\newtheorem{lemma}{Lemma}

\usepackage[ruled,linesnumbered]{algorithm2e}
\usepackage{amsmath}

\usepackage{cite}

\renewcommand{\algorithmcfname}{Algorithm}

\def\BibTeX{{\rm B\kern-.05em{\sc i\kern-.025em b}\kern-.08em
		T\kern-.1667em\lower.7ex\hbox{E}\kern-.125emX}}

\begin{document}

	\title{
Performance Analysis for Downlink Transmission in Multi-Connectivity Cellular V2X Networks	
 }

\author{
	\IEEEauthorblockN{Luofang Jiao,\textit{ Member, IEEE}, Jiwei Zhao, \textit{Member, IEEE}, Yunting Xu, \textit{Member, IEEE}, Tianqi Zhang, \textit{Member, IEEE}, Haibo Zhou, \textit{Senior Member, IEEE} and Dongmei Zhao, \textit{Senior Member, IEEE}}\\

   	\thanks{
   				This work is supported in part by the  National Natural Science Foundation Original Exploration Project of China under Grant 62250004, the National Natural Science Foundation of China under Grant 62271244, the Natural
   			Science Fund for Distinguished Young Scholars of Jiangsu Province under Grant BK20220067, the High-level Innovation and Entrepreneurship Talent Introduction Program Team of Jiangsu Province under Grant JSSCTD202202.
   		
   		L. Jiao, J. Zhao, Y. Xu, T, Zhang, and H. Zhou (Corresponding author) are with the School of Electronic
   	Science and Engineering, Nanjing University, Nanjing 210023, China
   	(e-mail: luofang\_jiao@smail.nju.edu.cn; jw\_zhao@smail.nju.edu.cn; yuntingxu@smail.nju.edu.cn; tianqizhang@smail.nju.edu.cn; haibozhou@nju.edu.cn).

D. Zhao is with the Department of Electrical and Computer Engineering, McMaster University, Hamilton, ON L8S 4L, Canada (e-mail: dzhao@mcmaster.ca).

Copyright (c) 2023 IEEE. Personal use of this material is permitted. However, permission to use this material for any other purposes must be obtained from the IEEE by sending a request to pubs-permissions@ieee.org.
}
}

\maketitle
\begin{abstract}

With the ever-increasing number of connected vehicles in the fifth-generation mobile communication networks (5G) and beyond 5G (B5G), ensuring the reliability and high-speed demand of cellular vehicle-to-everything (C-V2X) communication in scenarios where vehicles are moving at high speeds poses a significant challenge.
Recently, multi-connectivity technology has become a promising network access paradigm for improving network performance and reliability for C-V2X in the 5G and B5G era. To this end, this paper proposes an analytical framework for the performance of downlink in multi-connectivity C-V2X networks. Specifically, by modeling the vehicles and base stations as one-dimensional Poisson point processes, we first derive and analyze the joint distance distribution of multi-connectivity. Then through leveraging the tools of stochastic geometry, the coverage probability and spectral efficiency are obtained based on the previous results for general multi-connectivity cases in C-V2X. Additionally, we evaluate the effect of path loss exponent and the density of downlink base station on system performance indicators. We demonstrate through extensive Monte Carlo simulations that multi-connectivity technology can effectively enhance network performance in C-V2X. Our findings have important implications for the research and application of multi-connectivity C-V2X in the 5G and B5G era.

\end{abstract}
\IEEEpeerreviewmaketitle

\begin{IEEEkeywords}
C-V2X, multi-connectivity, coverage probability, spectral efficiency, stochastic geometry.
\end{IEEEkeywords}

\section{Introduction}

\IEEEPARstart W{ith} the evolution of the fifth-generation mobile communication networks (5G) and beyond 5G (B5G), reliable and high-performance wireless communication systems have become essential to fully exploit the potential of intelligent transportation systems (ITS) \cite{wolf2018reliable, pupiales2021multi, weedage2023impact}. Cellular vehicle-to-everything (C-V2X) has emerged as a promising technology that utilizes the existing cellular network infrastructure and spectrum to provide efficient and reliable communication between vehicles (V2V), as well as between vehicles and other network entities such as base stations (BS), roadside units, pedestrians, and cloud servers (V2I) \cite{xu2021leveraging}. This technology aims to enhance the efficiency and safety of vehicular traffic while enabling various applications such as collision avoidance, traffic management, cooperative driving, platooning, and autonomous driving, that require high-speed and low-latency communication in ITS \cite{chen2017vehicle, chen2017capacity}. However,
as the number of connected vehicles continues to increase, C-V2X is rapidly developing towards ultra-dense, whcih poses challenges such as interference management, security, and energy efficiency, and require further research and development to enable reliable and efficient vehicular communication.
C-V2X also faces significant challenges due to the high-speed mobility and dynamic topology of vehicles, which may cause rapid fluctuations in the quality of wireless links, frequent handovers, and increased signaling overhead. These challenges is severely impacting the communication performance and influence the quality of experience (QoE) and quality of service (QoS) of ITS applications.

In recent years, multi-connectivity technology has attracted significant attention to address these aforementioned challenges \cite{ lu2022personalized}. Multi-connectivity enables a vehicle to establish multiple simultaneous connections with different BSs using various radio access technologies (RATs), access points (APs), or channels, thus taking advantage of the diversity and availability of wireless resources in the cellular network \cite{weedage2023impact}.
The multi-connectivity technology, as compared to the traditional point-to-point communication, offers substantial advantages in terms of various communication performance metrics, e.g., enhanced reliability\cite{rabitsch2020utilizing}, improved coverage \cite{wu2020performance}, and enabled seamless mobility and handover\cite{tesema2016evaluation}.
Through accessing multiple BSs, vehicles can switch between available connections without disrupting ongoing communication, providing uninterrupted connectivity even when moving across network boundaries or transiting under different coverage of networks. In additions, based on the specific requirements of the application or user preferences, multi-connectivity is capable of offering flexibility for choosing the most suitable network connections and providing better spectral efficiency by dynamic adaptation to changing network conditions\cite{diez2018lasr}. Thus, applying multi-connectivity to C-V2X is of remarkable significance for improving communication performance and makes it more suitable for ITS applications.

In the context of multi-connectivity in C-V2X, coverage probability and spectral efficiency are two important performance metrics for evaluating wireless communication system performance \cite{wu2020performance}.
However, establishing a comprehensive analytical framework with regard to these performance metrics is challenging and holds significant academic importance.
 following a spatial point process \cite{chen2017performance, chetlur2018coverage}.
System modeling and analytical performance evaluation based on stochastic geometry have proven to be a powerful method for monitoring the effects of important system parameters as well as optimizing system configurations, all without the need for computationally expensive and resource-intensive computer simulations \cite{andrews2011tractable}.
Moreover, there is a research gap in the analysis of uplink and downlink performance for multi-connectivity C-V2X communication. While most of the existing studies primarily focused on the uplink transmission and performance optimization \cite{wu2020performance, qian2017non}, the downlink transmission based on an analytical framework has not been sufficiently explored. This is a significant limitation since downlink transmission plays a crucial role in supporting various ITS applications that rely on receiving downlink information from infrastructures. Therefore, conducting in-depth studies on the downlink performance of multi-connectivity C-V2X communication is substantial for filling this existing gap and ensuring a holistic analysis of the system's capabilities for supporting diverse ITS applications.

\textcolor{black}{ Therefore, this paper considers multi-connectivity as an effective solution to resolve the challenges faced by C-V2X communication, aiming to enhance the communication performance for ITS applications.
A feasible analytical framework for downlink transmission in multi-connectivity C-V2X networks is proposed by modeling the vehicles and downlink base stations (DBSs) as one-dimensional (1-D) Poisson point processes (PPPs), the tools of stochastic geometry are used to derive crucial performance indicators, including joint distance distribution, coverage probability, and spectral efficiency. The key contributions of this paper are summarized below: }
\begin{itemize}
	\item \textcolor{black}{We present a novel multi-connectivity performance analytical framework for C-V2X, which enables the evaluation of network performance in the 5G/B5G era. This framework provides a foundation for further research and  potential performance improvement of multi-connectivity technology in C-V2X systems.}

	\item \textcolor{black}{We derive precise expressions of coverage probability and spectral efficiency for general multi-connectivity cases in C-V2X based on the joint distance distribution. We also provide important insights into the design and optimization of C-V2X networks by analyzing the effect of path loss exponent and  DBS density on system performance indicators.}
	
	\item \textcolor{black}{We conduct comprehensive Monte Carlo simulations to confirm the effectiveness of the presented multi-connectivity performance analytical framework, which shows that multi-connectivity technology can significantly improve network performance in C-V2X. This finding has important implications for the practical applications of multi-connectivity C-V2X in the 5G/B5G era.}

\end{itemize}

The subsequent sections of this paper are structured as follows.
We briefly introduce the existing research works related to our
work in Section \uppercase\expandafter{\romannumeral2}. Section \uppercase\expandafter{\romannumeral3} presents the proposed framework for analyzing multi-connectivity performance.
Section \uppercase\expandafter{\romannumeral4} conducts a performance analysis of the system, including the joint distance distribution, coverage probability and spectral efficiency.
In Section \uppercase\expandafter{\romannumeral5}, the simulation setup and results obtained from extensive Monte Carlo simulations are presented, providing verification of the proposed framework and evaluation of the system performance.
Section \uppercase\expandafter{\romannumeral6} presents the concluding remarks of this paper.

\section{Related work}

\textcolor{black}{The adoption of C-V2X has emerged as a critical network paradigm for enabling vehicular communication with other vehicles and the infrastructure, offering diversified safety and efficiency applications for ITS \cite{jiao2022spectral}. However, despite of its significant potential, the implementation of C-V2X communication is confronted with numerous challenges, such as high mobility, dynamic topology, heterogeneous network, and stringent QoS requirements \cite{chen2017performance}.  In high-speed C-V2X scenarios, single connectivity with just one base station frequently leads to handover issues \cite{wu2020performance, lu2022personalized}, resulting in a rapid decline in communication speed and reliability, thus no longer meeting C-V2X's QoS requirements.}
\textcolor{black}{In recent years, multi-connectivity has been considered to be a promising technology to tackle these challenges in C-V2X through enhancing reliability, reducing latency, and boosting overall network performance.}

\textcolor{black}{With multiple BSs access, multi-connectivity supports seamless mobility and handover between different BS coverage. Exploiting  simultaneous connections, multi-connectivity offers considerable availability for the improvement of spectral efficiency \cite{xu2023federated}.
A number of studies have investigated the potential benefits of applying multi-connectivity in C-V2X and wireless networks.
Numerous studies have delved into the potential advantages of implementing multi-connectivity within C-V2X and wireless networks. These investigations cover various aspects, ranging from resource optimization in C-V2X multi-connectivity to performance analyses in wireless networks.}

\textcolor{black}{
In the context of C-V2X, some studies concentrate on optimizing communication resources. Rabitsch \textit{et al.} \cite{rabitsch2020utilizing} explored multi-connectivity algorithms tailored to meet the stringent requirements for communication availability and latency in V2I networks. Lu \textit{et al.} \cite{lu2022personalized} introduced a novel approach to reduce duplication rates in DBSs in fully-decoupled C-V2X networks. They achieved this by formulating and solving optimization problems using Lyapunov stochastic optimization techniques to help vehicles select access BSs for multi-connectivity and optimize bandwidth resources to meet user communication requirements. Kousaridas \textit{et al.} \cite{kousaridas2019multi} analyzed multi-connectivity management in a Manhattan model for V2X communication.}

\textcolor{black}{
Other studies aim to evaluate the performance of multi-connectivity in both wireless networks and C-V2X scenarios. Moltchanov \textit{et al.} \cite{moltchanov2018upper} provided a closed-form upper bound on the probability density function (PDF) for multi-connectivity, shedding light on its statistical characteristics. Weedage \textit{et al.} \cite{weedage2023impact} scrutinized the downlink performance of multi-connectivity in wireless networks. Wu \textit{et al.} \cite{wu2020performance} proposed a multi-connectivity scheme for uplink C-V2X communications, deriving precise expressions for the outage probability using stochastic geometry tools.}

Numerous performance metrics have been employed in research works that explore the application of multi-connectivity technology in cellular communication scenarios.
In 5G and beyond 5G networks, Sylla \textit{et al.} \cite{sylla2022multi} provided a comparable cellular communication analysis for multi-connectivity. Pupiales \textit{et al.}  \cite{pupiales2021multi} focused on the multi-connectivity architectures and protocols for 5G network and they described the different network entities and protocol layers involved in multi-connectivity, such as multi-connectivity coordinator, multi-connectivity agent, multi-connectivity manager and packet data. Petrov \textit{et al.} \cite{petrov2017dynamic} studied the dynamic characteristics of multi-connectivity technology, whereas Giordani \textit{et al.} \cite{giordani2016multi} investigated its application in 5G mmWave cellular networks.

In this paper, we mainly focus on the performance indicator of coverage probability and spectral efficiency based on distance distribution.
Coverage probability and spectral efficiency are two of the most important metrics that hold significance for evaluating the wireless networks. Firstly, coverage probability determines the reliability of C-V2X communication in different geographical areas and network densities \cite{chetlur2018coverage}. By leveraging multi-connectivity technology, the coverage probability can be improved and the risk of communication interruptions can be reduced. With multiple connections simultaneously receiving and transmitting data, even if one connection encounters interruption issues, the others can maintain communication, thereby enhancing overall coverage probability.
Secondly, spectral efficiency is extremely crucial for the transmission capacity of C-V2X. C-V2X communication involves handling a substantial amount of traffic-related information, including vehicle sensor data and traffic management instructions \cite{sattar2019spectral}. Additional spectrum resources can be utilized in parallel or through multiplexing, thus improving spectral efficiency to support higher data transmission rates and faster response times through multi-connectivity \cite{shafie2020multi}.
Further improvements in coverage probability and spectral efficiency can be achieved by optimizing load balancing and resource allocation among the connections. Research in this area is of paramount importance to achieve efficient and reliable C-V2X communication, providing a more robust and efficient foundation for critical applications such as real-time vehicle communication, traffic management, and vehicular safety in ITS.

\textcolor{black}{To obtain the exact analytical expression of performance metrics for multi-connectivity in C-V2X, leveraging the tools of stochastic geometry is regarded as an efficient approach and it has been increasingly popular in recent years for the performance analysis in multi-connectivity scenarios.
For instance, Moltchanov \textit{et al.} \cite{moltchanov2018upper} were among the pioneers in deriving the PDF for multi-connectivity, laying essential groundwork for further investigations. Building upon this foundation, Kibria \textit{et al.} \cite{kibria2018stochastic} assessed the viability of employing dual connectivity and coordinated multiple points (CoMP) transmission in wireless communication systems, expanding the scope of multi-connectivity applications.}

\textcolor{black}{
Moreover, the utilization of stochastic geometry in various multi-connectivity scenarios has witnessed extensive exploration. Shafie \textit{et al.} \cite{shafie2020multi} explored multi-connectivity in indoor communication systems using ultra-wideband terahertz (THz) technology, focusing on average ergodic capacity and connectivity likelihood. Chen \textit{et al.} \cite{chen2017performance} employed coordinated multipoint techniques to enhance spectral efficiency, while Giordani \textit{et al.} \cite{giordani2016multi} and  Kamble \textit{et al.} \cite{kamble2009efficient} aimed to optimize the Signal-to-Interference-Plus-Noise Ratio (SINR) and outage probability in single-frequency networks. Weedage \textit{et al.} \cite{weedage2023impact} delved into the analysis of channel capacity and outage probability in wireless networks' downlink scenarios.}

\textcolor{black}{\section{SYSTEM MODEL}}
\textcolor{black}{When analyzing the overall performance of a multi-connectivity C-V2X network, a specific vehicle is considered as a typical analysis object. Therefore, the multiple roads model can still be simplified and analyzed as a single road situation. Moreover, it has been proven that the single road model can effectively reflect the performance of multi-connectivity in C-V2X \cite{wu2020performance, lu2022personalized}.}
Therefore, to investigate the downlink multi-connectivity in C-V2X scenario, we introduce a simplified 1-D system model in this paper. A coordination scheme called Single Frequency Networks (SFN) as in \cite{simsek2019multiconnectivity} are leveraged for spectrum allocation. SFN enables the transmission of incoherent joint signals on the same radio resources in frequency and time, which requires BSs to coordinate when creating signals and to strictly synchronize their timing.
Our focus in this paper is on the intra-frequency multi-connectivity, which requires simultaneous transmission of multiple DBSs operating at the same carrier frequency to the same vehicle. This is an important issue to address in the C-V2X scenario, where high data rates and reliable communication are need for safety-critical applications.
The following of this section introduces the channel model, association policy, interference model, and performance metrics utilized in this study.
Table \ref{MAJOR SYMBOLS} lists the key symbols used throughout this paper.

\subsection{Modeling of C-V2X Network}

\begin{figure}[t]
	\centering

	\centerline{\includegraphics[width=1.0\hsize]{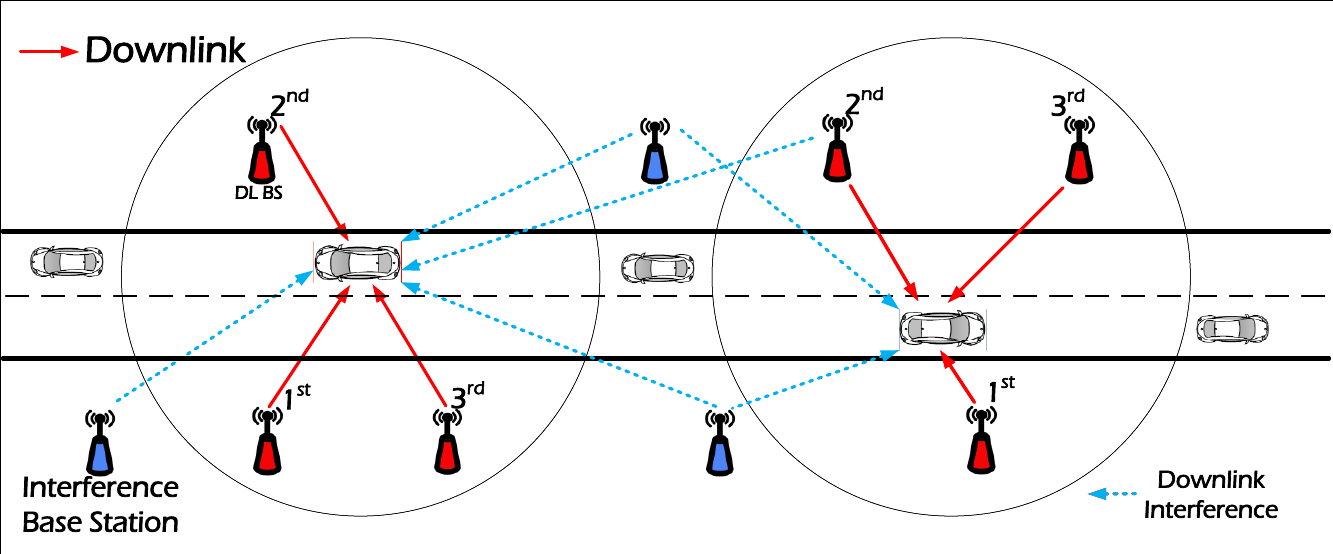}}
	\caption{An example of a practical 1-D scenario for downlink transmission in multi-connectivity C-V2X is illustrated. In this scenario, the target vehicle receives messages from the three closest DBSs, while transmissions from DBSs located beyond the collaboration distance can lead to interference to the target vehicle.}
	\label{1-d system-model}
\end{figure}

Fig. \ref{1-d system-model} shows the downlink multi-connectivity scenario in C-V2X networks. The vehicles are randomly distributed on an urban freeway segment, and the DBSs are densely distributed along the road. To simplify the analysis, we make the assumption that both the DBSs and vehicles utilize a single antenna, and denote the height difference between the antenna of the DBS and the vehicle as $h$.

\begin{table*}[h]
	\centering
	\color{black}\caption{A LIST OF MAJOR SYMBOLS}
	\begin{tabular}{c|c }	
		\hline
	 \label{MAJOR SYMBOLS}
		Notation & Description\\
		\hline
		 $\lambda$; $ x_{i} $& {The density of the vehicles and DBSs on a 1-D road;} {The distance of $ i $th nearest DBS to the typical vehicle. } \\
		$ P_d $; 	$ \alpha_{d}$ & {The transmit power of DBS;} {The downlink path loss exponent parameter.} \\
		$ g_d$; ${\chi_d} $&  {The channel gain between the DBS and vehicle;}  {The Nakagami-m fading gain.}\\
	$ \tau _{D} $	; 	$ \mu $ & {The spectral efficiency of downlink;}  {The mean of exponential function.}\\
		$\omega_d$; 	$\delta_d^2$ &{The mean of the logarithm of $\chi_d$;} {The variance of the logarithm of $\chi_d$.}\\
		${\varphi _V}, \varphi _D $ & {The Poisson point processes of vehicles and DBSs.}\\
		$\varphi_D^t $; 	$ t $& {The DBSs set after being executed random displacement; {The predetermined threshold $ t $ of coverage probability.} }\\
		$ \varphi _c $; 	$ \Theta _I^d $  & {The collaborative DBS set of multi-connectivity;} {The interference DBS set of multi-connectivity.}\\

		$ I_d $;	$ \sigma _d^2  $& {The received interference of the typical vehicle;} {The noise of channel.}\\

		$ \mathbb{E}\left ( \cdot  \right ) $ & {The expectation of a random variable.}\\
		$ \mathbb{P}\left ( \cdot  \right ) $ & {The probability of a random variable.}\\
		$ F\left ( \cdot  \right ) $ & {The cumulative distribution function of a random variable.}\\
		$ f\left ( \cdot  \right ) $ & {The probability density function of a random variable.}\\
		$ \zeta _{I}\left( \cdot  \right) $ & {The Laplace transform of interference $ I $.}\\
		$ \Gamma\left ( \cdot \right ) $ &{The gamma distribution function of a random variable.}\\
		\hline
	\end{tabular}
\end{table*}

For the tractability of the downlink performance analysis, we consider a 1-D scenario on a road including vehicles, DBSs, and interference DBSs as shown in Fig. \ref{1-d system-model}, as in \cite{wu2020performance, chen2017capacity}. From a statistical perspective, the spatial distributions of vehicles and DBSs conform to 1-D PPP distributions \cite{chetlur2018coverage}, and we use the 1-D PPPs $ {\varphi _V}, \varphi _D $ with density $ {\lambda _v}, \lambda _d$ to denote the locations of vehicles and DBSs on the road, respectively, where $ {\varphi _V}, \varphi _D $ can be expressed as
\begin{align}
\varphi_j\mathop {=}\limits^{\bigtriangleup}\left \{ x_{i,j}\in R^2: i\in \mathbb{N}_+ \right \},j=\left \{ V, D \right \}.\nonumber
\end{align}
All of the vehicles and DBSs are distributed along a road with length $ l $. As per Slivnyak's theorem, the distribution of point processes remains unchanged even after adding a node at the origin \cite{chiu2013stochastic}, and in order not to lose generality and eliminate segmentation due to boundary effects, we place the typical vehicle at the origin $ v_o=(0, 0) $, i.e. which represents the center of the road \cite{wu2020performance}.

In relation to the formation of virtual cells, we assume that each vehicle is connected to the $ n $ nearest DBSs on a Euclidean plane. The 1-D distance between the typical vehicle $ v_o $ with the $ i $-th ($ i\le n $) DBS is $ r_i $, thus the actual distance $ x_i $ between the transmit antenna of DBS to receive antenna of the typical vehicle is
\begin{equation}
x_i=\sqrt{r_i^2+h^2}.
\end{equation}

We adopt a common power-law pathloss and Rayleigh fading model with a decay rate of $ x ^{-\alpha_d } $, where $ x $ denotes the distance between the DBS and the typical vehicle. The downlink pathloss exponent parameter is denoted as $ \alpha_d $ $(\alpha_d>2)$. $ g_d $ is used to denote the power gain of Rayleigh fading and it is modeled by an exponential distribution with a mean of $1/\mu$. Therefore, we have $ g_d \sim exp\left( \mu \right) $. The distribution function of $ g_d $ is
\begin{align}
 f\left( g_d\right ) =\mu  e^{-\mu g_d}.
\end{align}
\textcolor{black}{Furthermore, we use random variable $ {\chi _{d}} $ to model the effects of shadowing between the DBS and the typical vehicle in the downlink, and $ {\chi _{d}} $  follows a log-normal distribution given by $ 10log_{10}\chi_{d}\sim \aleph \left ( \omega _{d},\delta _{d}^{2} \right ) $, where $\omega_d$ represents the mean of the logarithm of $\chi_d$ (i.e., the geometric mean of $\chi_d$), while $\delta_d^2$ represents the variance of the logarithm of $\chi_d$ \cite{abdulqader2015performance}.}
Hence, the received signal power of the typical vehicle from the $ i $-th DBS in the downlink is \cite{chetlur2019coverage}
\begin{equation}
	P_{r,v}(x_i) = {{P_d}{g_{d}}{\chi _{d}}{{ x_i }^{ - {\alpha _d}}},} ~{i \in  {\varphi_D}},
\end{equation}
where $ P_d $ is the transmitting power of the DBS and assumed to be the same for all DBSs.

\subsection{Association policy}
The typical vehicle is assumed to be connected to $ n $ nearest DBSs by measuring all the receiving power from the nearby DBSs, finding the DBSs with the maximum receiving power (MRP) in turn \cite{wu2020performance}.
Since the received power $ P_{r,v} $ is not exponentially distributed for the modeling of the shadow fading \cite{jiao2022spectral}, the lemma of random displacement theorem is considered to solve this issue \cite{xu2016wireless}.
Thus, $ P_{r,v}(x_i)={P_d}{g_{d}}{\chi _{d}}{{ x_i }^{ - {\alpha _d}}}$ can be transformed to $ P_{r,v}(y_i)={P_d}{g_{d}}{{ y_i }^{ - {\alpha_d }}}  $, where $ y_i={\chi_{d}^{-\frac{1}{\alpha_d }}}x_i $.
The 1-D PPP transformed converges to a 1-D homogeneous PPP and the intensity $ \lambda_d $ is transformed to $ E\left [ {\chi_{d}^{-\frac{1}{\alpha_d }}}  \right ]\lambda_d $, and the intensity of the 2-D PPP is $ E\left [ {\chi_{d}^{-\frac{2}{\alpha_d }}}  \right ]\lambda_d $ after executing the procedure of random displacement \cite{chetlur2018coverage}. Specifically, $ E\left [ {\chi_{d}^{-\frac{1}{\alpha_d }}}  \right ]\lambda $ can be calculated as
\begin{align}
\mathbb{E}\left[ {{\chi_d ^{ - \frac{1}{\alpha_d}}}} \right]\lambda_d  = \exp \left( {\frac{{\omega_d \ln 10}}{{10\alpha_d }} + \frac{1}{2}{{\left( {\frac{{\sigma_d \ln 10}}{{10\alpha_d }}} \right)}^2}} \right)\lambda_d.
\end{align}
Then we use 1-D PPP $ \varphi _{_D}^t $ to denote the transformed set of DBS and $ \lambda_{D} = E\left [ {\chi_{}^{-\frac{1}{\alpha_d }}}  \right ]\lambda_d  $ denotes the transformed DBS intensity. To facilitate performance analysis in the following sections, we use the symbol $ \varphi_D^{t,d} $ to denote the set of distances between the DBSs and the typical vehicle,
\begin{align}
\varphi_D^{t,d}=\{x_1,x_2, \dots, x_{i}\}, i\in \mathbb{N}_+,
\end{align}
where $ x_i $ denotes the distance between the typical vehicle $ v_o $ and the $ i $-th nearest DBS $ \in \varphi_D^{t} $.
Thus the candidate serving DBSs is changed to the  $ n $ nearest DBSs $ \in  \varphi _D^t $ in turn, and this can be expressed as
	\begin{align} \label{asso}
x_i = \mathop {\arg \max }\limits_{x_i \in {\varphi _D^{t,d}\backslash\varphi _c^d  }} x_i^{ - {\alpha _d}}, i>m
\end{align}
where we use $ \varphi_c^d=\{x_1,x_2, \dots, x_{m}\} $ to denote the set of the distances between the connected collaborative DBSs and the typical vehcle $ v_o $ at the origin, $ m $ is the number of DBSs that the typical vehicle has already connected to, and $ x_{i},{ i \in \{m+1, m+2, \cdots\}}$ denotes the distance between the DBS outside  $\varphi _c^d$ and the typical vehicle. Let $ \varphi _c $ to denote the set of the connected collaborative DBSs. This means that to expand the set $ \varphi _c $, we need to find the nearest DBS among DBSs in $\varphi _D^{t}\backslash\varphi _c  $. 

\subsection{Interference}

 In the collaboration DBSs set $ \varphi _c $, all DBSs will transmit the control and data signals simultaneously on the same subband \cite{lu2022personalized}.
Since the signal components of the DBSs are within the cyclic prefix, the resulting multi-connectivity SINR experienced by the typical vehicle $v_o$ in the downlink is defined as follows:

\begin{align} \label{SINR}
S\!I\!N\!{R_D} = \frac{{\sum\limits_{i \in \varphi _c} {{P_d}{g_d}x_i^{ - {\alpha _d}}} }}{{{I_D} + \sigma _d^2}},
\end{align}
\textcolor{black}{where $ \sum\limits_{i \in \varphi _c} {{P_d}{g_d}x_i^{ - {\alpha _d}}} $ represents the sum of received signal power from the DBSs in $ \varphi_c $. We use $ \sigma_d^2 $ to denote the power of the additive white Gaussian noise (AWGN) \cite{andrews2011tractable}. $ I_D $ is the power of aggregate interference from the DBSs outside of $ \varphi_c $ }and $ I_D $ can be expressed as
\begin{align} \label{interfer}
{I_D} = \sum\limits_{i \in \left\{ {\varphi _{_D}^t\backslash {\varphi _c}} \right\}} {{P_d}{g_d}x_i^{ - {\alpha _d}}}.
\end{align}

\vspace{0.5cm}
\subsection{Performance Metrics}

In order to enable advanced C-V2X applications such as automated driving applications and stream media \cite{xu2021leveraging, jiao2022spectral}, it is crucial to ensure that the downlink transmission is both reliable and capable of transmitting data at a high rate. This is important not only from the perspective of a single vehicle but also from the perspective of the whole C-V2X network. To this end, this paper conducts an analytical evaluation of two performance metrics, i.e. coverage probability and spectral efficiency as follows.
\begin{itemize}
	\item The coverage probability of the typical vehicle $ v_o $ in downlink, is defined as the probability that the received SINR outperforms a predetermined threshold $ t $ \cite{yang2015coverage}. It can be expressed as
	\begin{align}
	{\mathbb{P}_{{\mathop{ cov}} }}\left( t \right) = \mathbb{P}\left( {S\!I\!N\!{R_D} >t } \right).
	\end{align}
	It can also be calculated as the proportion of vehicles that have the received $ S\!I\!N\!{R_D} $ above a threshold $ t $, i.e., establish a successful connection with the DBSs in $ \varphi _c $, among all vehicles in the simulation scenario. Since the cumulative distribution function (CDF) of $ S\!I\!N\!{R_D} $ is 	$ {\mathbb{P}_{{\mathop{ cov}} }}\left( t \right) = \mathbb{P}\left( {S\!I\!N\!{R_D} <t } \right)  $, the coverage probability can also be expressed as the complementary cumulative distribution function (CCDF) of the $ S\!I\!N\!{R_D} $ at the typical vehicle from the DBSs.
	
	\item The spectral efficiency of the typical vehicle $ v_o $ is the amount of data transmitted per unit of bandwidth \cite{jia2017downlink}. According to the Shannon Theory, the spectral efficiency of the downlink is
	\begin{align}
		\tau_D = \mathbb{E}\left[ {\ln \left( {1 + S\!I\!N\!{R_D}} \right)} \right],
	\end{align}
	where $ \mathbb{E}(\cdot ) $ is the expectation function. The spectral efficiency describes the likelihood of a wireless communication system achieving a specific information amount within a certain time period and space range during actual use \cite{sattar2019spectral}. It can help evaluate the performance of C-V2X in a multi-connectivity environment and determine whether system optimization or adjustments are needed \cite{jia2017downlink}.
	
\end{itemize}

\vspace{0.5cm}
\section{Performance Analysis}

We first derive the expression for the joint distance distribution from $ x_1 $ to $ x_n $ in this section.
To optimize system configurations without the need for time-consuming computer simulations, we leverage the stochastic geometry. Specifically, by using the tools provided by stochastic geometry,
 we utilize the results obtained from previous sections to derive the coverage probability and spectral efficiency of C-V2X in a multi-connectivity scenario.

\textcolor{black}{\subsection{The joint distance distribution of the typical vehicle to $ n $ service DBSs} }

Since the typical vehicle is connected to the $ n $ nearest DBSs in multi-connectivity, no other DBSs are closer than distance $ x_n $. And it also means that all interference DBSs are farther than $ x_n $. The above definition can be expressed by $ f\left( {{x_1},{x_2}, \cdots, {x_n}} \right) $, and we call it joint distance distribution for $ {{x_1},{x_2}, \cdots, {x_n}} $.
\begin{lemma} \label{joint distance distribution}
The joint distance distribution of the typical vehicle to its service DBSs in set $ \varphi{_c} $  from $ x_1 $ to $ x_n $ is
		\begin{align} \label{f1n}
f\left( {{x_1},{x_2}, \cdots, {x_n}} \right) = {\left( {2{\lambda _D}} \right)^n}{e^{ - 2{\lambda _D}{x_n}}},
	\end{align}
	where $ x_n $ denotes the distance between the typical vehicle and the $ n $-th closest DBS in $ \varphi_c $.
\end{lemma}
\begin{IEEEproof}
The null probability of a PPP in an area $ A $ is $e^{-\lambda A}$, where $A=2\lambda x$ in 1-D PPP and $ A= \pi x^2 $ in 2-D PPP, thus the CCDF of $ x_1 $ is \cite{andrews2011tractable}
\begin{align}
\mathbb{P}\left [ x >x_1 \right ] &= \mathbb{P}\left [ \text{no~DBS~closer~than~} x_1 \right ] \nonumber\\
&=e^{-2\lambda_D x_1}.
\end{align}
\textcolor{black}{Because the $ C\!D\!F=1-C\!C\!D\!F $, the CDF of $ x_1 $ is}
	\begin{align}
F\left( x_1 \right) = 1 - {e^{ - 2{\lambda _D}x_1}}.
\end{align}
Since the PDF $f\left ( x  \right ) =\frac{\partial F\left (x  \right )}{\partial x} $ \cite{chiu2013stochastic}, the PDF of $ x_1 $ is
\begin{align}
f\left( x_1 \right) = 2{\lambda _D}{e^{ - 2{\lambda _D}x_1}}.
\end{align}

According to the definition of Section 3.3 in \cite{moltchanov2012distance},
let $ f\left( {{x_2}|{x_1}} \right) $ denote the probability that the 2nd closest DBS is at $ x_2 $ given that the closest one is at the distance of $ x_1 $. Thus the probability of having no DBSs between the distances $ x_1 $ and $ x_2 $ can be calculated as follows
  	\begin{align} \label{f1}
  	f\left( {{x_2}|{x_1}} \right) = 2{\lambda _D}{e^{ - 2{\lambda _D}\left( {{x_2} - {x_1}} \right)}}.	\end{align}

According to the conditional probability Bayes theorem \cite{berrar2018bayes}, $ f\left( {{x_2},{x_1}} \right) $ denotes the joint distance distribution to the two nearest distances, i.e., the probability of having at least one point in $ x_2+\triangle x $, where $ \triangle x $ is an infinitesimal quantity, is
	\begin{align}\label{f12}
	f\left( {{x_1},{x_2}} \right) = f\left( {{x_2}|{x_1}} \right)f\left( {{x_1}} \right) = {\left( {2{\lambda _D}} \right)^2}{e^{ - 2{\lambda _D}{x_2}}}.\end{align}
By following the similar procedures in Eq. \eqref{f1} and Eq. \eqref{f12}, the joint distance distribution $ f\left( {{x_1},{x_2} \cdots {x_n}} \right) $ from $ x_1 $ to $ x_n $ is
		\begin{align} \label{jpdf}
	f\left( {{x_1},{x_2}, \cdots, {x_n}} \right) = {\left( {2{\lambda _D}} \right)^n}{e^{ - 2{\lambda _D}{x_n}}}.
	\end{align}
\end{IEEEproof}

To compare the joint distance distribution  $f\left( {{x_1},{x_2}, \cdots, {x_n}} \right) $  and the PDF of $ x_n $, we provide the PDF of $ x_n $ in Eq. \eqref{pdfxn} as
	\begin{align} \label{pdfxn}
	f\left( {{x_n}} \right) = \frac{{{{\left( {2{\lambda _b}{x_n}} \right)}^n}}}{{{x_n}\Gamma \left( n \right)}}{e^{ - 2{\lambda _b}{x_n}}}, \end{align}
	where $ \Gamma ( n )= (n-1)! $ when $n$ is a positive integer.

\subsection{Coverage Probability}

A general expression for the coverage probability of multi-connectivity in C-V2X is calculated in this subsection.
\begin{theorem} \label{CP}
	A vehicle is considered to be within the coverage area if its $ S\!I\!N\!R_D $ value from the nearest base station exceeds a certain threshold value $ t $. On the other hand, if the $ S\!I\!N\!R_D $ falls below $ t $, the vehicle is dropped from the network.
Thus, the coverage probability of downlink for multi-connectivity C-V2X is
\begin{align}
&\mathbb{P}\left( {{S\!I\!N\!{R_D}} > {{t}}} \right)\nonumber\\
&= \int_{0 < {x_1} < {x_2} <  \cdots  < {x_m} < \infty } {{\zeta _{I_D^{}}}\left( j \right)\exp \left( { - \frac{{\mu t\sigma _d^{{2}}}}{{\sum\limits_{{{i = 1}}}^{{m}} {{P_d}x_i^{{{ - }}{\alpha _d}}} }}} \right) \times } \nonumber\\
&f\left( {{x_1},{x_2}, \cdots ,{x_m}} \right)d{x_1}d{x_2} \cdots d{x_m},
\end{align}
 where $ j = \frac{\mu t}{{\sum\limits_{{{i = 1}}}^{{m}} {{P_d}x_i^{{{ - }}{\alpha _d}}} }} $, $ m $ is the number of cooperating DBSs in the cooperative set, $ \zeta _{I_D^{}}\left( j \right)  $ is the Laplace transform of random variable interference $ I_D $ evaluated at $ j $ and $ \zeta _{I_D^{}}\left( j \right)  $ is
 \begin{align}
 &{\zeta _{{I_D}}}\left( j \right) =
  \exp \left[ { - 2{\lambda _D}\int_{x_m}^\infty  {1 - \frac{\mu}{{j{P_d}x_i^{ - {\alpha _d}} + \mu}}d{x_i}} } \right].
 \end{align}
\end{theorem}

\begin{IEEEproof}
The proof of coverage probability in the downlink is
\begin{align}
&{\mathbb{P}}\left( {{{S\!I\!N\!}}{{{R}}_D} > {{t}}} \right)\nonumber\\
&{{ \mathop  = \limits^{\left( a \right)} \mathbb{P}}}\left( {\frac{{\sum\limits_{{{i = 1}}}^{{m}} {{P_d}{g_d}x_i^{{{ - }}{\alpha _d}}} }}{{{I_D}{{ + }}\sigma _d^{{2}}}} > {{t}}} \right)\nonumber\\
&= \mathbb{P}\left( {{g_d} > \frac{{t\left( {{I_D}{{ + }}\sigma _d^{{2}}} \right)}}{{\sum\limits_{{{i = 1}}}^m {{P_D}x_i^{{{ - }}{\alpha _d}}} }}} \right)\nonumber\\
&\mathop  = \limits^{\left( b \right)} {\mathbb{E}_{{x_i},{I_D}}}\left[ {\exp \left( { - \frac{{\mu t\left( {{I_D}{{ + }}\sigma _d^{{2}}} \right)}}{{\sum\limits_{{{i = 1}}}^{{m}} {{P_d}x_i^{{{ - }}{\alpha _d}}} }}} \right)} \right]\nonumber\\
&\mathop  = \limits^{\left( c \right)} {\mathbb{E}_{{x_i}}}\left[ {\exp \left( { - \frac{{\mu t\sigma _d^{{2}}}}{{\sum\limits_{{{i = 1}}}^{{m}} {{P_d}x_i^{{{ - }}{\alpha _d}}} }}} \right){\zeta _{I_D^{}}}\left( j \right)} \right]\nonumber\\
&= \int_{0 < {x_1} < {x_2} <  \cdots  < {x_m} < \infty } {{\zeta _{I_D^{}}}\left( j \right)\exp \left( { - \frac{{\mu t\sigma _d^{{2}}}}{{\sum\limits_{{{i = 1}}}^{{m}} {{P_d}x_i^{{{ - }}{\alpha _d}}} }}} \right) \times } \nonumber\\
&f\left( {{x_1},{x_2}, \cdots ,{x_m}} \right)d{x_1}d{x_2} \cdots d{x_m},\end{align}
where (a) is obtained by substituting the expression of $ S\!I\!N\!R_D $ in Eq. \eqref{SINR}.
(b) is obtained by finding the CCDF of $g_d$ which is exponentially distributed with parameter $\mu$.
$ \zeta _{I_D^{}}\left( j \right)  $ is the Laplace transform of interference $ I_D $ in (c), and $ j = \frac{ \mu t}{{\sum\limits_{{{i = 1}}}^{{m}} {{P_d}x_i^{{{ - }}{\alpha _d}}} }} $.
Based on the definition of the Laplace transform, the derivation of $ \zeta _{I_D^{}}\left( j \right)$ is
\begin{align} \label{laplacet}
&{\zeta _{{I_D}}}\left( j \right) = {\mathbb{E}_{{I_D}}}\left[ {{e^{ - j{I_D}}}} \right]\nonumber\\
&\mathop  = \limits^{(a)} {\mathbb{E}_{{I_D}}}\left[ {\exp \left( { - j\sum\limits_{i \in \varphi _D^t\backslash \varphi _c }^{} {{P_d}{g_d}x_i^{ - {\alpha _d}}} } \right)} \right]\nonumber\\
&\mathop  = \limits^{\left( b\right)} {\mathbb{E}_{\Theta _I^d,\left\{ {{g_d}} \right\}}}\left[ {\prod\limits_{i \in \Theta _I^d} {{e^{ - j{P_d}{g_d}x_i^{ - {\alpha _d}}}}} } \right]\nonumber\\
&\mathop  = \limits^{\left( c \right)} \exp \left[ { - 2{\lambda _D}\int_{{x_m}}^\infty  {1 - } } \right.\nonumber\\
&\left. {{\mathbb{E}_{{g_d}}}\left[ {\exp \left( { - j{P_d}{g_d}x_i^{ - {\alpha _d}}} \right)} \right]d{x_i}} \right]\nonumber\\
&\mathop  = \limits^{(d)} \exp \left[ { - 2{\lambda _D}\int_{x_m}^\infty  {1 - \frac{\mu}{{j{P_d}x_i^{ - {\alpha _d}} + \mu}}d{x_i}} } \right],
\end{align}
where we use $ {\Theta _I^d}= \varphi _D^t\backslash \varphi _c $ to denote the interference DBSs, interference $ I_D $ can be obtained in Eq. \eqref{interfer}.
(b) is obtained by finding the CCDF of $g_d$ which is exponentially distributed with parameter $\mu$.
(c) is derived from the probability generating functional (PGFL) of the PPP \cite{stoyan2013stochastic}, i.e.,
\begin{align}
\mathbb{E}\left ( \prod f\left ( x \right )  \right ) =exp\left ( -\lambda \int_{R^2}\left ( 1-f\left ( x \right )  \right ) dx  \right ).
\end{align}
$ \mathbb{E}_{{g_d}}\left[ {\exp \left( { - j{P_d}{g_d}x_i^{ - {\alpha _d}}} \right)} \right] $ in (d) can be derived as
\begin{align}
&{\mathbb{E}_{{g_d}}}\left[ {\exp \left( { - j{P_d}{g_d}x_i^{ - {\alpha _d}}} \right)} \right]\nonumber\\
&= \int_0^\infty  {{e^{ - j{P_d}{g_d}x_i^{ - {\alpha _d}}}}{\mu e^{ - \mu{g_d}}}} d{g_d}\nonumber\\
&=  - \frac{{{\mu e^{ - j\left( {{P_d}x_i^{ - {\alpha _d}} + \mu} \right){g_d}}}}}{{j{P_d}x_i^{ - {\alpha _d}} + \mu}}\left| {_0^\infty } \right.\nonumber\\
&= \frac{\mu}{{j{P_d}x_i^{ - {\alpha _d}} + \mu}}.
\end{align}
Since the farthest cooperation DBS is at a distance of $ x_m $, the integration limits are from $ x_m $ to $ \infty  $ in (d).

\end{IEEEproof}

\subsection{Spectral efficiency}

This subsection derives the expression of spectral efficiency for the downlink by using the tools of stochastic geometry for C-V2X in multi-connectivity.
We computed the spectral efficiency in units of $ nats/s/Hz $ ($ 1~bit = ln(2) =0.693~nats $) for the typical vehicle.

\begin{theorem} \label{SE} The spectral efficiency of the downlink in multi-connectivity C-V2X is
	\begin{align}
	\tau_D&	= \int_{0 < {x_1} < {x_2}< \cdots  < {x_m} < \infty }  f{\left( {{x_1},{x_2}, \cdots ,{x_m}} \right)}\times\nonumber\\
	&	\mathbb{E } \left[ {\ln \left( {1 + S\!I\!N\!R_D}\right)} \right]d{x_1}d{x_2} \cdots d{x_m},
\end{align}
where $ m $ is the number of cooperating DBSs. $ \mathbb{E } \left[ {\ln \left( {1 + S\!I\!N\!R_D} \right)} \right] $ is
	\begin{align}
&	\mathbb{E } \left[ {\ln \left( {1 + S\!I\!N\!{R_D}} \right)} \right]=\nonumber\\
&\int_{t > 0} {P\left[ \ln \left( {1 + \frac{{\sum\limits_{i = 1}^m {{P_d}{g_d}x_i^{ - {\alpha _d}}} }}{{{I_D} + \sigma _d^2}}} \right)>t \right]}dt,
\end{align}
where
\begin{align}
&\mathbb{P}\left[ {\ln \left( {1 + \frac{{\sum\limits_{i = 1}^m {{P_d}{g_d}x_i^{ - {\alpha _d}}} }}{{{I_D} + \sigma _d^2}}} \right)} > t \right]\nonumber\\
&\mathop  = \limits^{} \exp \left( { - \frac{{\beta \sigma _d^{{2}}}}{{\sum\limits_{{{i = 1}}}^{{m}} {{P_d}x_i^{{{ - }}{\alpha _d}}} }}} \right){\zeta _{I_D^{}}}\left( j \right),
{}\end{align}
where $ \beta  = \mu \left( {{e^t} - 1} \right)   $ and $ j = \frac{\beta }{{\sum\limits_{{{i = 1}}}^{{m}} {{P_d}x_i^{{{ - }}{\alpha _d}}} }} $. $ {\zeta _{I_D^{}}}\left( j \right) $ is the Laplace transform of interference $ I_D $, and $ {\zeta _{I_D^{}}}\left( j \right) $ is the same as in Eq. \eqref{laplacet},
\begin{align}
{\zeta _{I_D^{}}}\left( {{j_{}}} \right) = \exp \left[ { - 2{\lambda _D}\int_{{x_m}}^\infty  {\left( {1 - \frac{\mu}{{jP_dx_i^{ - {\alpha _d}} + \mu}}} \right)d{x_i}} } \right].
\end{align}

\end{theorem}
\begin{IEEEproof}
The proof of spectral efficiency of downlink is
	\begin{align} \label{sed}
\tau_D &= \mathbb{E } \left[ {\ln \left( {1 + S\!I\!N\!{R_D}} \right)} \right]\nonumber\\
&	= \int_{0 < {x_1} < {x_2}< \cdots  < {x_m} < \infty }  f{\left( {{x_1},{x_2}, \cdots ,{x_m}} \right)}\times\nonumber\\
&	\mathbb{E } \left[ {\ln \left( {1 + \frac{{\sum\limits_{i = 1}^m {{P_d}{g_d}x_i^{ - {\alpha _d}}} }}{{{I_D} + \sigma _d^2}}} \right)} \right]d{x_1}d{x_2} \cdots d{x_m}\nonumber\\
&{{ \mathop  = \limits^{(a)}}}\int_{0 < {x_1} < {x_2} <  \cdots  < {x_m} < \infty } {f\left( {{x_1},{x_2}, \cdots ,{x_m}} \right)}  \times\nonumber \\
&\int_{t > 0} {\mathbb{P}\left[ \ln \left( {1 + \frac{{\sum\limits_{i = 1}^m {{P_d}{g_d}x_i^{ - {\alpha _d}}} }}{{{I_D} + \sigma _d^2}}} \right)>t \right]}dtd{x_1}d{x_2} \cdots d{x_m},
\end{align}
where $ t $ is the predetermined threshold.
As a positive random variable $ X $ is considered, it follows that $ \mathbb{E}\left ( X \right ) $ can be calculated as $ \int_{0}^{\infty }\mathbb{P}(X>t)dt $ \cite{andrews2011tractable},
\textcolor{black}{thus the $ \mathbb{E } \left[ {\ln \left( {1 + S\!I\!N\!R_D} \right)} \right] $ can be calculated in (a).}
Furthermore, $ {\mathbb{P}\left[ \ln \left( {1 + {{\sum\limits_{i = 1}^m {{P_d}{g_d}x_i^{ - {\alpha _d}}} }}/({{{I_D} + \sigma _d^2}}}) \right) >t\right]} $ is
\begin{align}
&\mathbb{P}\left[ {\ln \left( {1 + \frac{{\sum\limits_{i = 1}^m {{P_d}{g_d}x_i^{ - {\alpha _d}}} }}{{{I_D} + \sigma _d^2}}} \right)} > t \right]\nonumber\\
&\mathop  = \limits^{\left( a \right)} \mathbb{P}\left( {\frac{{\sum\limits_{i = 1}^m {{P_d}{g_d}x_i^{ - {\alpha _d}}} }}{{I_D^{} + \sigma _d^2}} > {e^t} - 1} \right)\nonumber\\
&\mathop  = \limits^{\left( b \right)} {\mathbb{E}_{{g_d}}}\left[ {{g_d} > \frac{{\left( {{e^t} - 1} \right)\left( {I_D^{} + \sigma _d^2} \right)}}{{\sum\limits_{{{i = 1}}}^{{m}} {{P_d}x_i^{{{ - }}{\alpha _d}}} }}} \right]\nonumber\\
&\mathop  = \limits^{(c)} {\mathbb{E}_{{I_D}}}\left[ {\exp \left( { - \frac{{\mu \left( {{e^t} - 1} \right)\left( {{I_{{D}}}{{ + }}\sigma _d^{{2}}} \right)}}{{\sum\limits_{{{i = 1}}}^{{m}} {{P_d}x_i^{{{ - }}{\alpha _d}}} }}} \right)} \right]\nonumber\\
&\mathop  = \limits^{(d)} \exp \left( { - \frac{{\beta \sigma _d^{{2}}}}{{\sum\limits_{{{i = 1}}}^{{m}} {{P_d}x_i^{{{ - }}{\alpha _d}}} }}} \right){\zeta _{I_D^{}}}\left( j \right),
{}\end{align}
where (a) first solves the logarithm, then calculate the expectation of the channel gain $ g_d $ in (b), and $ g_d $ follows the exponential distribution with mean $ 1/\mu $ in (c). Since some variables have nothing to do with $ I_D $, they can be treated as constants and remain unchanged in (d). For the simplicity of the formula, we use $ \beta  = \mu \left( {{e^t} - 1} \right)   $ and $ j = \frac{\beta }{{\sum\limits_{{{i = 1}}}^{{m}} {{P_d}x_i^{{{ - }}{\alpha _d}}} }} $. The Laplace transform $ \zeta _{I_D} $ is the same with Eq. \eqref{laplacet} and is omitted here.

\end{IEEEproof}

\begin{figure}[t]
	\centering
	\centerline{\includegraphics[width=1.1\hsize]{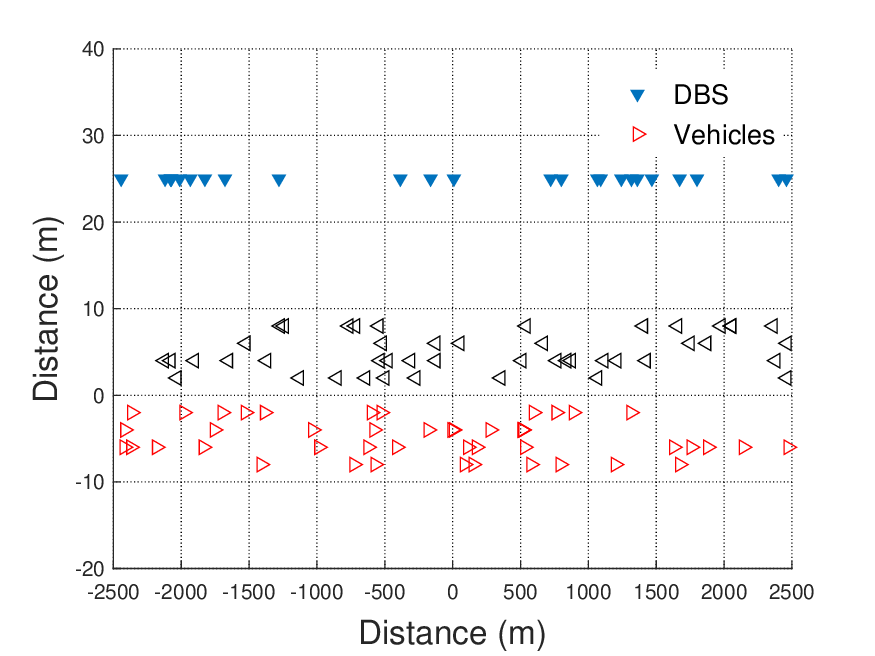}}
	\caption{Simulation scenario of multi-connectivity in C-V2X.}
	\label{simfig}
\end{figure}

\textcolor{black}{\subsection{Special case: Single-connectivity} }

\textcolor{black}{In order to compare the performance with multi-connectivity in C-V2X, this subsection focuses on the calculation of the coverage probability and spectral efficiency in a cellular single-connectivity scenario, which represents the most basic approach.} In this scenario, the typical vehicle associates with the cellular base station (CBS) whit the MRP.

As the single-connectivity is a special case of multi-connectivity,  we model the similar channel model as in multi-connectivity scenario, and use $\lambda_{C}$ to denote the transformed intensity $ \lambda_{c} $ of CBS $ \varrho  _{_C}^t $ after executing the procedures of random displacement, and $\lambda_{C} > \lambda_{D}$.
As the CBS are distributed along the road following a 1-D PPP, the PDF of distance distribution is
\begin{align}
f\left( x \right) = 2{\lambda _C}{e^{ - 2{\lambda _C}x}},
\end{align}
where $ x $ is the distance between the nearest CBS and the typical vehicle.

The coverage probability of the downlink in cellular single-connectivity is
\begin{align}
&{\mathbb{P}_{{\mathop{ cov}} }}\left( {S\!I\!N\!{R_c^D} > t} \right)\nonumber\\
&= \int_0^\infty 2{\lambda _C} {{e^{ -\mu t\sigma _d^2{x^{{\alpha _d}}}/{P_d}}}} {e^{ - 2{\lambda _C}x}}{\zeta _{I_d^c}}\left( j \right)dx,
\end{align}
where $ j =\mu t{x^{{\alpha _d}}}/{P_d} $, the $ S\!I\!N\!R_c^D $ is
 \begin{equation}
S\!I\!N\!R_c^D= \frac{{{P_d}{g_d}x_i^{{{ - }}{\alpha _d}}}}{{I_d^c{{ + }}\sigma _d^{{2}}}},
 \end{equation}
 where the interference $ I_d^c $ is
  \begin{equation}
 {I_d^c} = \sum\limits_{i \in  {\varrho  _{_C}^t} } {{P_d}{g_d}x_i^{ - {\alpha _d}}}.
  \end{equation}
  The Laplace transform of $ I_d^c $ is
\begin{align}
{\zeta _{I_d^c}}\left( j \right) = \exp \left[ { - 2{\lambda _C}\int_{_x}^\infty  {1 - \frac{\mu}{{j{P_d}x_i^{ - {\alpha _d}} + \mu}}d{x_i}} } \right].
\end{align}

\begin{IEEEproof}
Given the similarity in the proof to that of Theorem \ref{CP}, we omit the specific steps here.
\end{IEEEproof}

The spectral efficiency of cellular single-connectivity for downlink is

\begin{align}
\tau _c^D = \int_0^\infty  {f\left( x \right)} E\left[ {\ln \left( {1 + S\!I\!N\!R_c^D} \right) > t} \right]dx,
\end{align}
where
\begin{align}
\mathbb{E}\left[ {\ln \left( {1 + S\!I\!N\!R_c^D} \right) > t} \right]= \int_0^\infty  {{e^{ - \frac{{\mu \left( {{e^t} - 1} \right){x^{{\alpha _d}}}\sigma _d^2}}{{{P_d}}}}}{\zeta _{I_d^c}}\left( j \right)} dt,
\end{align}
where $ j =\mu ({e^t} - 1){x^{{\alpha _d}}}/{P_d} $, the $ {\zeta _{I_d^c}}\left( j \right)$ is
\begin{align}
&{\zeta _{I_d^c}}\left( j \right) =\nonumber \\
&~\exp \left[ { - 2{\lambda _D}\int_{_x}^\infty  {\left( {1 - \frac{\mu}{{\left( {{e^t} - 1} \right){x^{{\alpha _d}}}x_i^{ - {\alpha _d}} + \mu}}} \right)d{x_i}} } \right]S
\end{align}
\begin{IEEEproof}
The proof of spectral efficiency of downlink for cellular single-connectivity is similar to Theorem \ref{SE}, the specific steps are omitted here.
\end{IEEEproof}

\renewcommand{\algorithmcfname}{Simulation}
\begin{algorithm} [htbp]
	\color{black}	\caption{Simulation for multi-connectivity in C-V2X} 
	\label{alg-sim} 
	\KwIn{simulation number $n$, road length $ l $, threshold $ t $, DBS density $ \lambda_d $, vehicle density $ \lambda_v $;} 
	\KwOut{ Coverage probability {$ C\!P $}, spectral efficiency $ \tau_{D} $;} 
	Initialize  ${\bf \tau_o}\leftarrow{\bf 0}_{n \times \lambda_v l}$ , ${\bf P}\leftarrow{\bf 0}_{n \times \lambda_v l} $, ${  C\!P_n }\leftarrow{\bf 0}_{1 \times n}$, ${  \tau_n }\leftarrow{\bf 0}_{1 \times n}$  \\ 
	\For{$i=1;i \le n;i++$} 
	{ 
		Generate the locations $ \bf{m} $, $ \bf{V} $ of DBSs and vehicles following 1-D PPP, respectively;\\
		{ 
			\For{$v=1;v \le \lambda_v l;v++$} 	
			{
			 Select collaborative DBSs according to Eq. (\ref{asso});\\			
				Calculate $ S\!I\!N\!R_D $ of DL according to Eq. (\ref{SINR});\\
				$ {\bf{\tau_o}}( i ,v) = ln(1+ S\!I\!N\!R_D )$, ${\bf P}( i ,v ) =  S\!I\!N\!R_D $;
			}
		}
		$C\!P_n(i) = \sum_{i=1}^{\lambda_vl} ({\bf P} (i,:)>t)/(\lambda_v l)$;\\
		${\tau_n}(i)$ $= \sum_{i=1}^{\lambda_vl} $$ \bf{\tau_o} $($ i $,:) /($ \lambda_v l $);\\
		
	}
	Return  ${\bf{\tau}}_D = \sum_{i=1}^{n} $$\tau_n(i)/n  $, ${C\!P} = \sum_{i=1}^{n} $$C\!P_n(i)/n $;
\end{algorithm}

\section{NUMERICAL and Simulation RESULTS}
A two-tier communication scenario on a straight urban freeway is considered in this section. The length of the freeway is set as 30 km. The specific simulation scenario is shown in Fig. \ref{simfig}. We first verify the proposed theoretical derivation in previous sections over 10,000 Monte Carlo simulations of the DBSs and vehicles following 1-D PPPs.
\textcolor{black}{The detailed steps of the simulation are in Simulation \ref{alg-sim}.}
 We use `Cellu 1', `Conn 2', and `Conn 3' to abbreviate single-connectivity, dual-connectivity, and triple-connectivity, respectively, in the legends of the figures. According to \cite{3gpp.36.331, 3gpp.36.819, elshaer2016downlink},
Table \ref{table1} summarizes the system simulation parameters employed in this paper.

\begin{figure*}[htp]
	\subfigure[a]{
		\label{pdf1}
		\begin{minipage}[t] {0.33\linewidth}
			\centering
			\includegraphics[width=1\hsize]{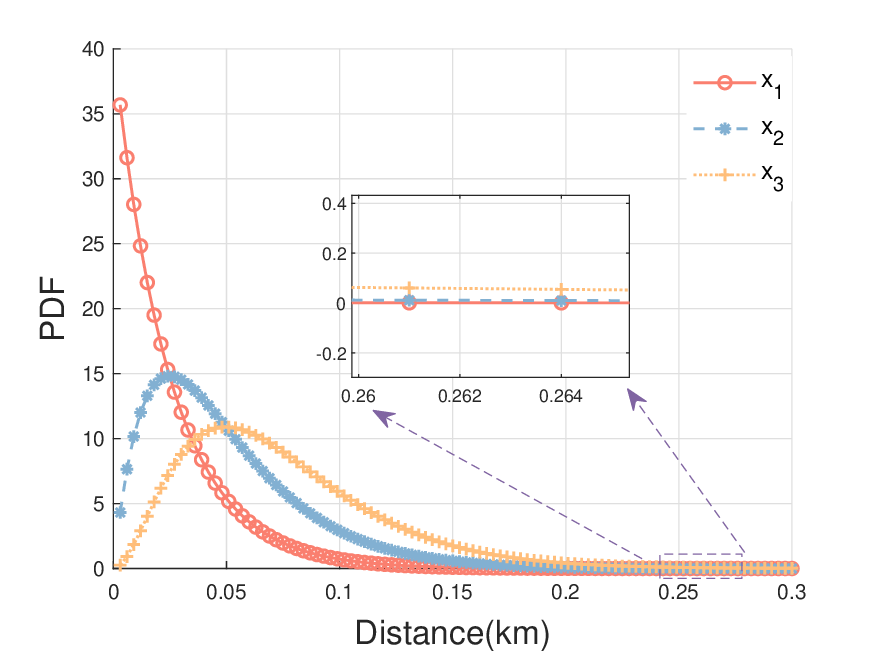}
		\end{minipage} }
	\subfigure[b]{		
		\label{pdf123}
		\begin{minipage}[t] {0.33\linewidth}
			\centering
			\includegraphics[width=1\hsize]{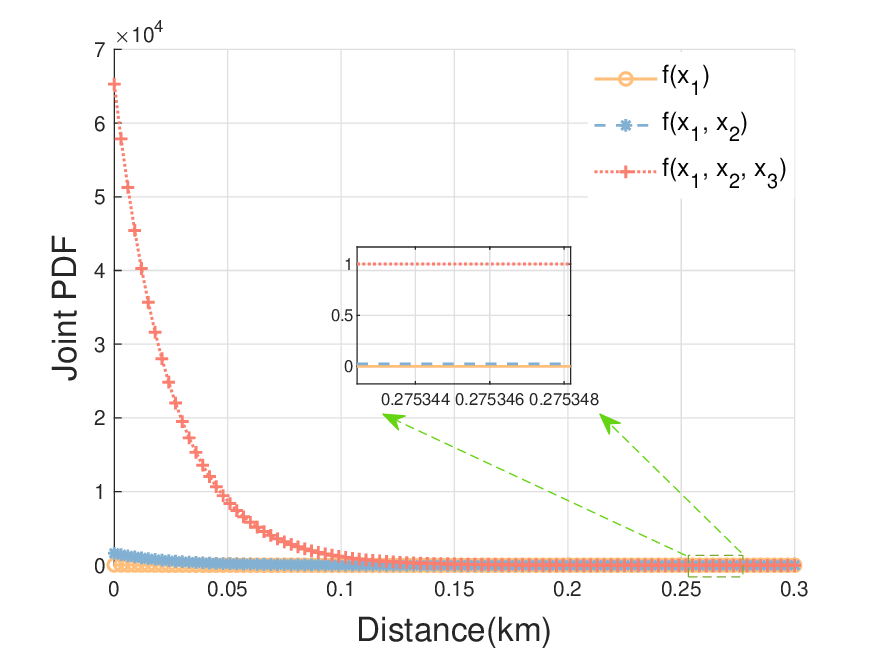}
		\end{minipage}	}
	\subfigure[c]{
		\label{pdf23}
		\begin{minipage}[t] {0.33\linewidth}
			\centering
			\includegraphics[width=1\hsize]{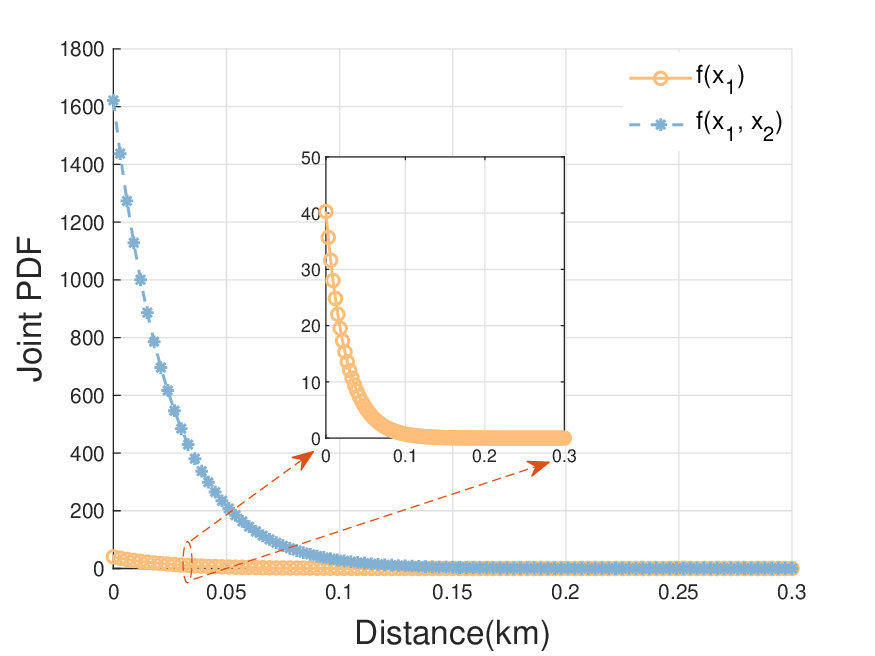}
		\end{minipage}	}	
	\caption{The distance distributions $ f(x_i) $ and joint distance distribution $ f\left( {{x_1},{x_2}, \cdots ,{x_m}} \right)$ under different distances. (a) Distance distribution for nearest distances $ x_1, x_2$ and $x_3 $. (b) The joint distance distribution of $ f(x_1), f(x_1, x_2) $ and $ f(x_1, x_2, x_3) $. (c) Since $ f(x_1) $ and $ f(x_1, x_2) $ is  much smaller than $ f(x_1, x_2, x_3) $, the two functions are highlighted here. }
	\label{alpha-lambda}
\end{figure*}

\begin{table}[h]
	\centering
\color{black}	\caption{MAIN PARAMETERS}
	\begin{tabular}{l|l }
		\hline
		\label{table1}
		Channel Parameters  & Value \\
		\hline
		DBS transmitting power $P_{d}$ (dBm)& 23 \\	
		Pathloss exponent for downlink  $\alpha _{d}$ &$ 2.1\sim 6 $ \\
		Noise power $\sigma _d^2 $ (dBm)    &-96 \\
		Mean of log-normal shadowing gain (dB) &0 \\
		Std of shadowing gain for MBS (dB) & 2\\
			\hline
		Simulation parameters&Value\\
		\hline
	    The length of road (km) & 30\\
		The number of iteration&10,000\\
	    Density of vehicle on road $ \lambda_{v}$  (nodes/km)   &20  \\
		Density of DBS $ \lambda_{d}$  (nodes/km)   &$ 0.05\sim 5.7 $ \\	
		Threshold (dB) & $ 0\sim 40 $\\
		\hline
	\end{tabular}
\end{table}

\subsection{Joint distance distribution}

 Fig. \ref{pdf1} shows the distance distribution of $ x_1 $, $ x_2 $, and $ x_3 $. We can see that the peak is gradually moving away from the origin from $ x_1 $ to $ x_3 $. Fig. \ref{pdf123} and Fig. \ref{pdf23} depict the joint distance distribution for $ f(x_1, x_2) $ and $ f(x_1,x_2,x_3) $. We can see that the peak of $ f(x_1,x_2,x_3) $ is closest to the origin, followed by $ f(x_1, x_2) $, and the furthest is $ f(x_1) $. Compared with distance distributions in Fig. \ref{pdf1}, the peak of joint distance distribution has a huge boost.
 The closer the distance between the peak and the origin, the better the performance. It can be observed that in single-connectivity, $ f(x_1)  $ exhibits better performance.
 Compared to a single-connectivity, a greater number of DBSs connections in multi-connectivity lead to a more significant performance improvement.

\subsection{Coverage probability}
\begin{figure}[t]
	\centering
	\centerline{\includegraphics[width=1.1\hsize]{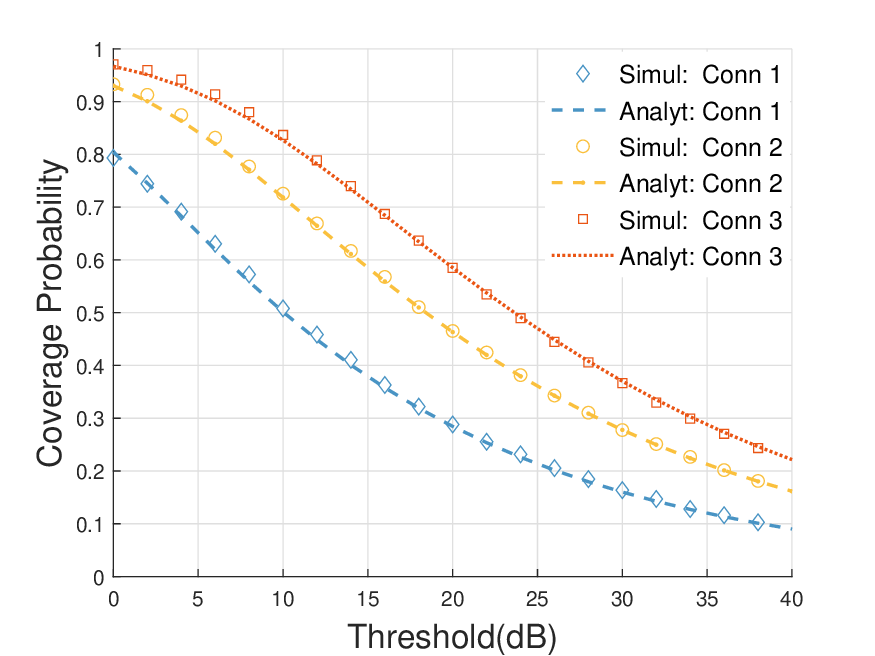}}
	\caption{Coverage probability variation with threshold $ t $ $ t \in [0,40] $ ($\lambda_d = 3~nodes/km, \lambda_c = 6~nodes/km$).}
	\label{CPT}
\end{figure}

The coverage probability variation of downlink with threshold $ t $ is illustrated in Fig. \ref{CPT}.
It is apparent that the simulation values closely match the theoretical values, which further verifies the validity of the theoretical derivation results.
The density of BS $ \lambda_{c} $ in single-connectivity is set as 3 nodes/km, and the density $\lambda_d$ of DBSs in multi-connectivity is set as 6 nodes/km. Though $\lambda_c > \lambda_{d}$, we can see that the dual-connectivity and triple-connectivity still have a greater coverage probability than single-connectivity. This suggests that multi-connectivity performs better than cellular single-connectivity in C-V2X and multi-connectivity enhances the coverage area of communications.

\begin{figure}[t]
	\centering
	\centerline{\includegraphics[width=.9\hsize]{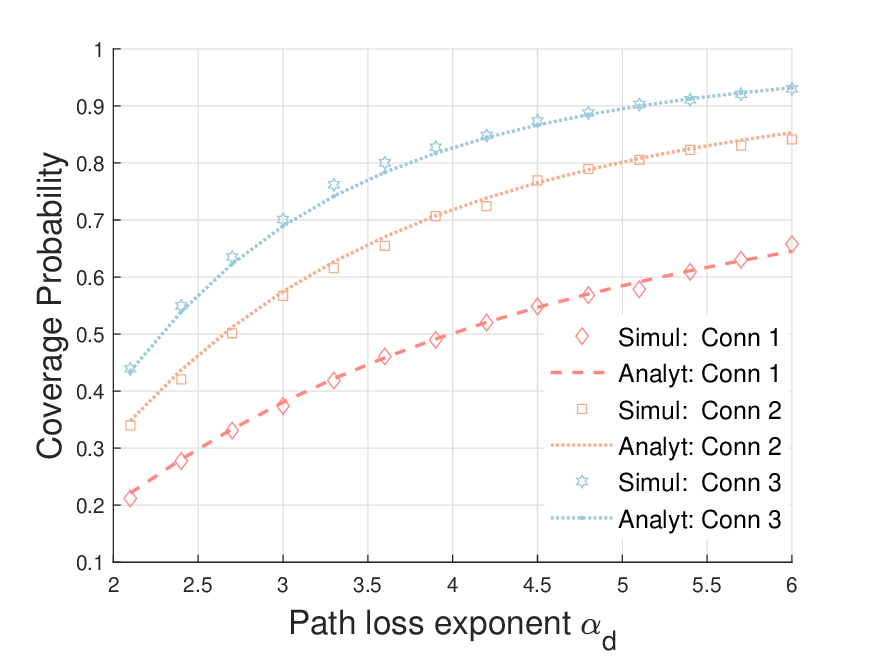}}
	\caption{The effect of path loss exponent on coverage probability( $ \alpha_d \in [2.1,  6]$).}
	\label{CPA}
\end{figure}

		Fig. \ref{CPA} illustrates the coverage probability as a function of path loss exponent $\alpha_{d}$. It can be seen that the Monte Carlo simulation data and analytical data fit well.
Considering the dense deployment of DBSs in the simulation, vehicles are in an interference-limited state. At this moment, the interference power will decrease in accordance with the increase of $\alpha_d$, which in turn lead to a promotion of SINR. Therefore, the coverage probability will be improved even if the channel gain decrease.

\begin{figure}[t]	
	\centering
	\begin{minipage} [htp] {0.48\textwidth}
		\includegraphics[width=.9\hsize]{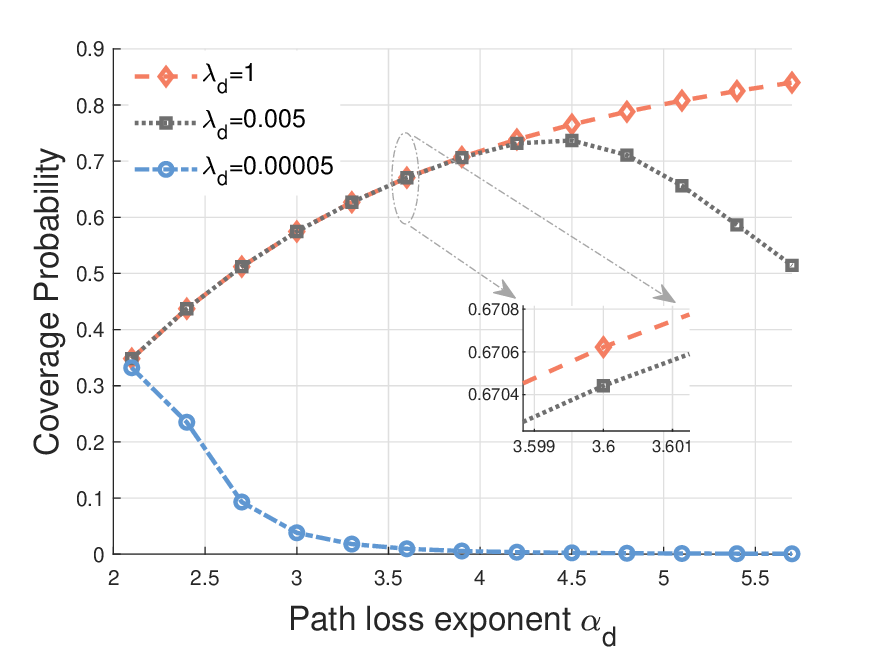}  		
		\caption{Coverage probability of dual-connectivity varies with different path loss exponent in different densities of DBSs $ \alpha_{d}~(2.1 \sim 5.7)$.}
		\label{cp di}
	\end{minipage}
\end{figure}

For a better investigation of the impact of path loss exponent $\alpha_{d}$ on the coverage probability under different densities of DBSs, we plot Fig. \ref{cp di} in a dual-connectivity scenario. As shown in Fig. \ref{cp di}, when the density of DBSs is in a dense deployment, the system is an interference-limited network. The distance between the signal DBSs $\in \varphi_c$ and the interference DBSs is close to the typical vehicle, so the increase of $\alpha_{d}$ leads to a greater impact on the interference signal power, resulting in an increase in the coverage probability. However, when the density is low enough, the system can be considered as a noise-limited network. Both the signal DBSs and the interference DBSs are far away from the typical vehicle, so the increase of $\alpha_{d}$ has a greater impact on the receiving signal power, leading to a continuous decrease in the coverage probability. When the density $\lambda_{d}$ is at an appropriate size, such as $\lambda_d$ $ = 0.005~ nodes/km $, the coverage probability first increases and then decreases with the increase of $\alpha_{d}$.

\begin{figure}[t]	
	\centering
	\begin{minipage} [htp] {0.48\textwidth}
		\includegraphics[width=.9\hsize]{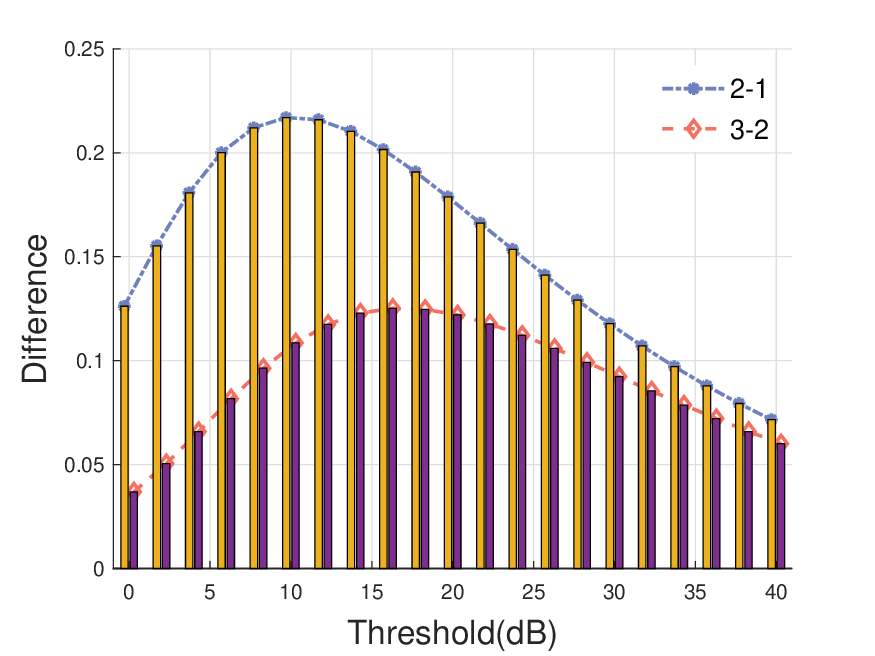}  	 
		\caption{Coverage probability differences variation with threshold $ t~(0 \sim 40 ~dB)$.}
		\label{cp diff}
	\end{minipage}
\end{figure}

The coverage probability of all cases decreases as the threshold $ t $ increases, while the difference between single-connectivity and multi-connectivity first increases and then decreases in Fig. \ref{cp diff}.
This is mainly because the coverage probability is respectively high and low at small and large thresholds, respectively. Only when the threshold value is in the middle range, the difference in coverage probability is large, and the advantage of applying multi-connectivity is also demonstrated.
It can be observed that increasing the number of connected DBSs does not result in a proportional increase in the coverage probability gain.
Hence, at a particular threshold, there exists a balance tradeoff between the number of corporation DBSs that are connected and the associated cost.

\begin{figure}[t]	
\centering
	\includegraphics[width=.9\hsize]{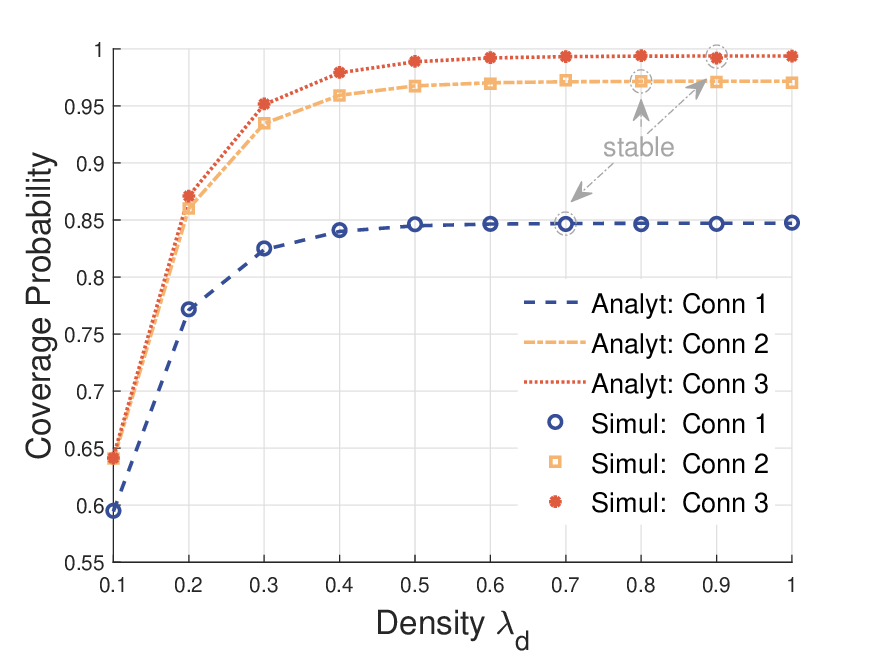}  	 
	\caption{Coverage probability variation with base station density $\lambda_d$ ($ 0.1 \sim 1 $ nodes/km).}
	\label{cpdensity}
\end{figure}
\begin{figure}[t]	
	\centering
	\begin{minipage} [t] {0.5\textwidth}
		\includegraphics[width=.9\hsize]{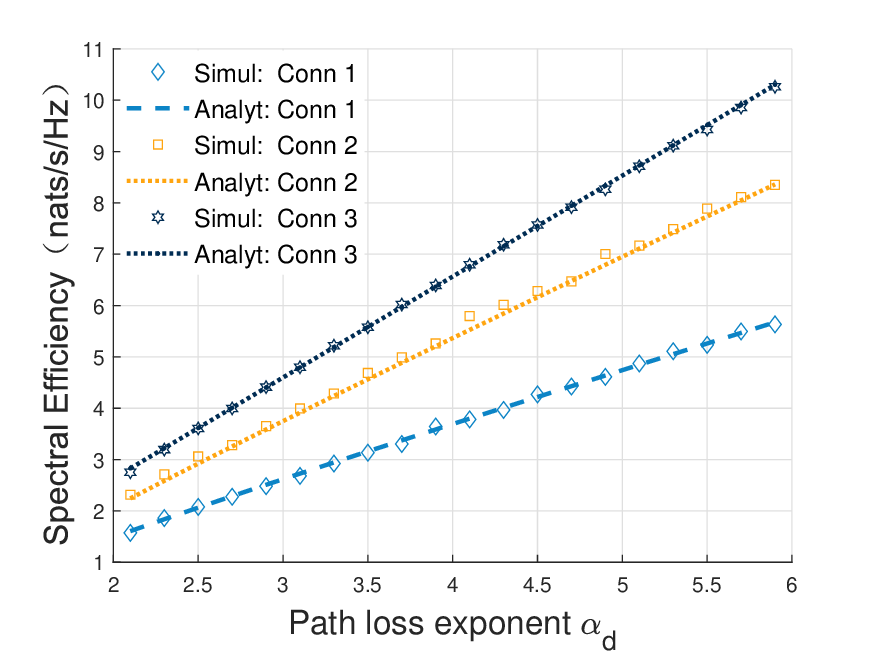}  	 
		\caption{Spectral efficiency variation with the path loss exponent $\alpha_{d}~(2.1 \sim 5.9)$.}
		\label{sea}
	\end{minipage}
\end{figure}
\begin{figure}[t]	
	\centering
	\begin{minipage} [t] {0.5\textwidth}
		\includegraphics[width=.9\hsize]{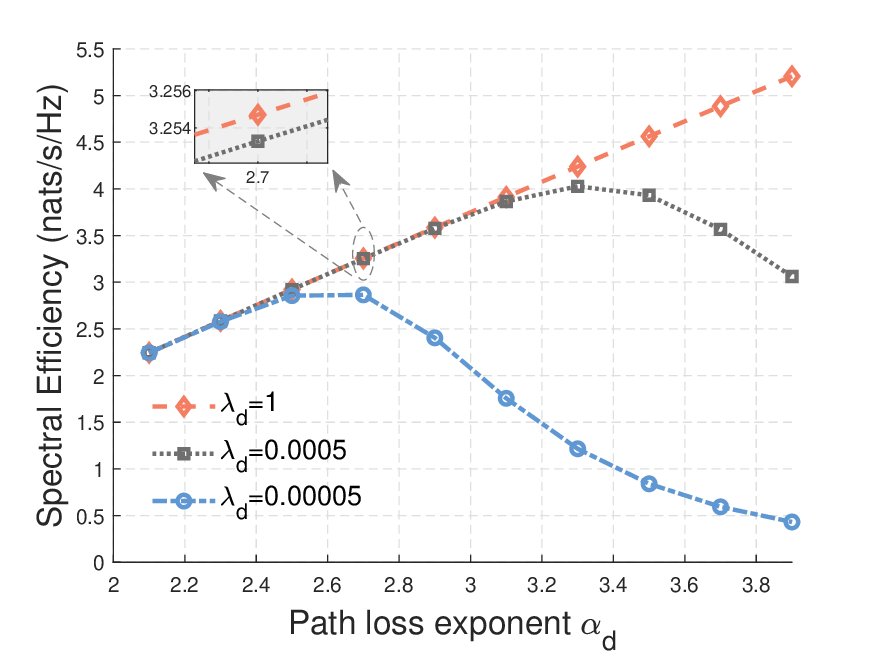}  	 
		\caption{Spectral efficiency of dual-connectivity varies with different path loss exponent in different densities of DBSs $ \alpha_{d}~(2.1 \sim 3.9)$.}
		\label{seaden}
	\end{minipage}
\end{figure}
\begin{figure}[t]	
\begin{minipage} [t] {0.5\textwidth}
	\includegraphics[width=.90\hsize]{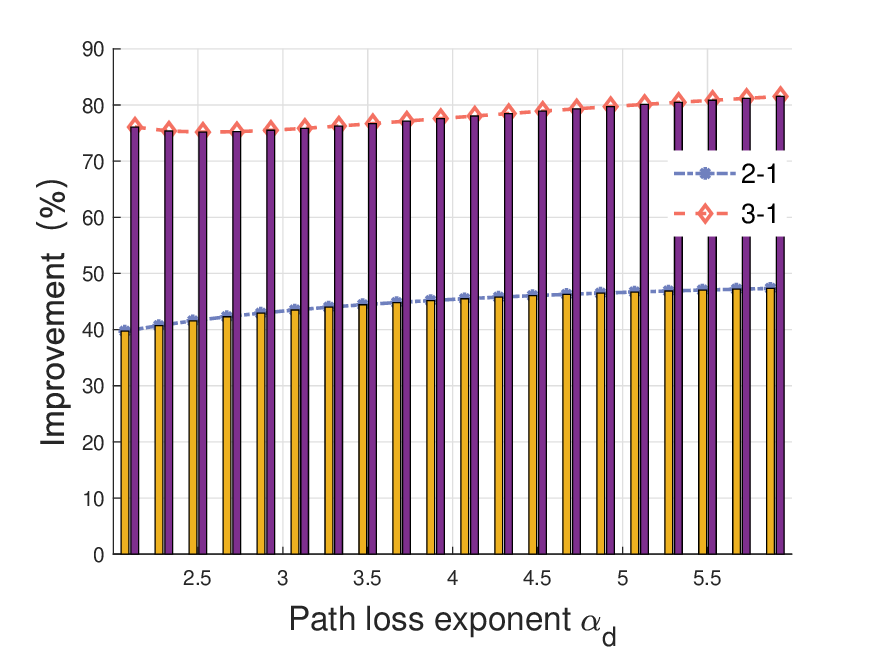}  	 
	\caption{Spectral efficiency percentage increase of multi-connectivity compared to single-connectivity.}
	\label{sediff}
\end{minipage}
\end{figure}
\begin{figure}[t]	
	\begin{minipage} [t] {0.5\textwidth}
		\includegraphics[width=.86\hsize]{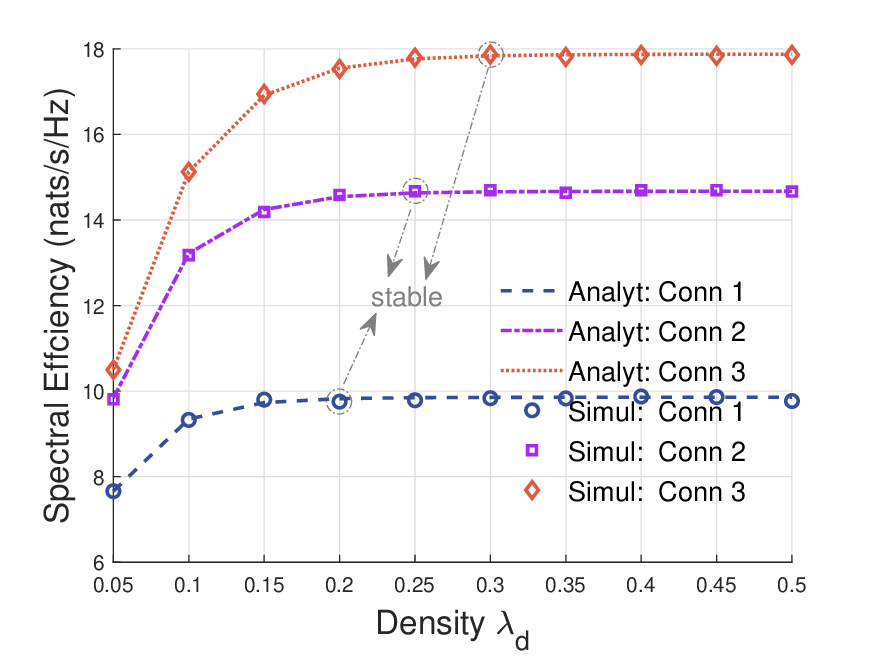}  	 
		\caption{Spectral efficiency as a function of base station density $\lambda_d$ ($0.05 \sim 0.5$ nodes/km ).}
		\label{sedensity}
	\end{minipage}	
\end{figure}

As shown in Fig. \ref{cpdensity},
the coverage probability varies with the density $\lambda_d $ of DBSs.
The coverage probability goes up first and then stays relatively constant for both cellular single-connectivity and multi-connectivity when the density $ \lambda_d $ increases.
It can also be seen that the growth rate of multi-connection is faster than that of single-connection, and it also reaches a stable point later. At the same time, it can be seen that the coverage probability does not increase significantly when comparing triple-connectivity to dual-connectivity. Therefore, increasing the number of cooperative BSs can improve communication coverage area, but it may also incur high costs.
By observing the horizontal axis, it can be noted that the change in DBS density $\lambda_d $ is relatively small in magnitude. This implies that the coverage probability is sensitive to changes in $\lambda_d $, and suggests that simply increasing $\lambda_d $ does not necessarily lead to an improvement in coverage probability.

\subsection{Spectral efficiency}

The spectral efficiency varies with the path loss exponent $\alpha_{d}$ is illustrated in Fig. \ref{sea}. The simulation data is represented by the points and the theoretical data is represented by the dashed line in Fig. \ref{sea}.
We can observe that the simulation data matches the theoretical data well, demonstrating the correctness of the theoretical derivation.
The increase of path loss exponent leads to a gradual increase in spectral efficiency and an approximately linear relationship.
The reason is that, as the dense deployment of DBSs, vehicles are in a interference-limited state, with the increase of path loss exponent $\alpha_{d}$, the interference signals weaken more than the signals transmitted by the cooperative BSs, resulting in an increase in SINR and consequently spectral efficiency.
It appears that there exists a close-to-linear relationship between the path loss exponent $\alpha_{d}$, mainly because after dividing the received signal power $ {{\sum\limits_{i \in \varphi_c } {{P_d}{g_d}x_i^{ - {\alpha _d}}} }} $ into the denominator of Eq. \eqref{sea}, since the thermal noise is much smaller than the received signal and thus $ \sigma _d^2/{{\sum\limits_{i \in \varphi_c } {{P_d}{g_d}x_i^{ - {\alpha _d}}} }}\approx 0 $, then after logarithmic calculation, it approximates to a linear relationship.
Fig. \ref{sedensity} depicts the spectral efficiency of dual-connectivity as a function of $\alpha_{d}$ in different densities of DBSs. Since the spectral efficiency is mainly affected by SINR, it can be seen that the trend of spectral efficiency is similar to that of coverage probability in Fig. \ref{cp di} when the DBSs layout changes from extremely dense to extremely sparse.

Fig. \ref{sediff} plots the spectral efficiency improvements between the multi-connectivity and single-connectivity.
It can be observed that the application of multi-connectivity technology greatly improves the spectral efficiency in C-V2X.
 The improvement achieved by dual-connectivity can reach up to 40\%, and that achieved by triple-connectivity can increase to more than 75\%.
With the increase of path loss exponent $\alpha_d$, the performance improvement of spectral efficiency does not increase significantly. This indicates that multi-connectivity technology has a stable performance gain.

Fig. \ref{sedensity} illustrates the spectral efficiency varies with the DBS density $\lambda_d$. With the increase of $\lambda_d$, the spectral efficiency first improves and then remains stable. Moreover, the improvement of multi-connectivity is greater than that of single-connectivity, and the stable point is also further back.
This means that multi-connectivity has a larger range of performance improvement.
Similar to Fig. \ref{cpdensity}, adding too many DBSs does not continuously improve spectral efficiency. Additionally, it can be observed that the stable point of spectral efficiency arrives earlier than the  stable point of coverage probability. Thus, when increasing the density of DBSs $\lambda_d$, it is necessary to comprehensively consider the demands between spectral efficiency and coverage probability.

\section{CONCLUSION}

This paper has demonstrated the potential of enhancing network performance in C-V2X by using the proposed multi-connectivity performance analysis framework.
 By analyzing performance indicators such as coverage probability and spectral efficiency in the downlink, this paper has provided insights into the effect of path loss exponent and the density of DBS on the system performance indicators. The extensive Monte Carlo simulations have effectively validated the proposed framework and demonstrated the effectiveness of multi-connectivity technology in enhancing the performance of C-V2X networks.
 The results of this paper have important implications for the research and practical applications of multi-connectivity C-V2X in the 5G and B5G era, and further investigations are warranted to explore the full potential of this technology for next-generation communication systems.

\bibliographystyle{IEEEtran}
\bibliography{references}

\begin{thebibliography}{10}
\providecommand{\url}[1]{#1}
\csname url@samestyle\endcsname
\providecommand{\newblock}{\relax}
\providecommand{\bibinfo}[2]{#2}
\providecommand{\BIBentrySTDinterwordspacing}{\spaceskip=0pt\relax}
\providecommand{\BIBentryALTinterwordstretchfactor}{4}
\providecommand{\BIBentryALTinterwordspacing}{\spaceskip=\fontdimen2\font plus
\BIBentryALTinterwordstretchfactor\fontdimen3\font minus
  \fontdimen4\font\relax}
\providecommand{\BIBforeignlanguage}[2]{{%
\expandafter\ifx\csname l@#1\endcsname\relax
\typeout{** WARNING: IEEEtran.bst: No hyphenation pattern has been}%
\typeout{** loaded for the language `#1'. Using the pattern for}%
\typeout{** the default language instead.}%
\else
\language=\csname l@#1\endcsname
\fi
#2}}
\providecommand{\BIBdecl}{\relax}
\BIBdecl

\bibitem{wolf2018reliable}
A.~Wolf, P.~Schulz, M.~D{\"o}rpinghaus, J.~C.~S. Santos~Filho, and G.~Fettweis,
  ``{How reliable and capable is multi-connectivity?}'' \emph{IEEE Transactions
  on Communications}, vol.~67, no.~2, pp. 1506--1520, 2018.

\bibitem{pupiales2021multi}
C.~Pupiales, D.~Laselva, Q.~De~Coninck, A.~Jain, and I.~Demirkol,
  ``{Multi-connectivity in mobile networks: Challenges and benefits},''
  \emph{IEEE Communications Magazine}, vol.~59, no.~11, pp. 116--122, 2021.

\bibitem{weedage2023impact}
L.~Weedage, C.~Stegehuis, and S.~Bayhan, ``{Impact of multi-connectivity on
  channel capacity and outage probability in wireless networks},'' \emph{IEEE
  Transactions on Vehicular Technology}, 2023.

\bibitem{xu2021leveraging}
Y.~Xu, H.~Zhou, T.~Ma, J.~Zhao, B.~Qian, and X.~Shen, ``Leveraging multiagent
  learning for automated vehicles scheduling at nonsignalized intersections,''
  \emph{IEEE Internet of Things Journal}, vol.~8, no.~14, pp. 11\,427--11\,439,
  2021.

\bibitem{chen2017vehicle}
S.~Chen, J.~Hu, Y.~Shi, Y.~Peng, J.~Fang, R.~Zhao, and L.~Zhao,
  ``{Vehicle-to-everything (V2X) services supported by LTE-based systems and
  5G},'' \emph{IEEE Communications Standards Magazine}, vol.~1, no.~2, pp.
  70--76, 2017.

\bibitem{chen2017capacity}
J.~Chen, G.~Mao, C.~Li, W.~Liang, and D.-g. Zhang, ``{Capacity of cooperative
  vehicular networks with infrastructure support: Multiuser case},'' \emph{IEEE
  Transactions on Vehicular Technology}, vol.~67, no.~2, pp. 1546--1560, 2017.

\bibitem{lu2022personalized}
Z.~Lu, T.~Zhang, X.~Ji, B.~Qian, L.~Jiao, and H.~Zhou, ``{Personalized Wireless
  Resource Allocation in Multi-Connectivity B5G C-V2X Networks},'' in
  \emph{2022 14th International Conference on Wireless Communications and
  Signal Processing (WCSP)}.\hskip 1em plus 0.5em minus 0.4em\relax IEEE, 2022,
  pp. 1--6.

\bibitem{rabitsch2020utilizing}
A.~Rabitsch, K.-J. Grinnemo, A.~Brunstrom, H.~Abrahamsson, F.~B. Abdesslem,
  S.~Alfredsson, and B.~Ahlgren, ``{Utilizing multi-connectivity to reduce
  latency and enhance availability for vehicle to infrastructure
  communication},'' \emph{IEEE Transactions on Mobile Computing}, vol.~21,
  no.~5, pp. 1874--1891, 2020.

\bibitem{wu2020performance}
P.~Wu, L.~Ding, Y.~Wang, L.~Li, H.~Zheng, and J.~Zhang, ``{Performance analysis
  for uplink transmission in user-centric ultra-dense V2I networks},''
  \emph{IEEE Transactions on Vehicular Technology}, vol.~69, no.~9, pp.
  9342--9355, 2020.

\bibitem{tesema2016evaluation}
F.~B. Tesema, A.~Awada, I.~Viering, M.~Simsek, and G.~Fettweis, ``{Evaluation
  of context-aware mobility robustness optimization and multi-connectivity in
  intra-frequency 5G ultra dense networks},'' \emph{IEEE Wireless
  Communications Letters}, vol.~5, no.~6, pp. 608--611, 2016.

\bibitem{diez2018lasr}
L.~Diez, A.~Garcia-Saavedra, V.~Valls, X.~Li, X.~Costa-Perez, and
  R.~Ag{\"u}ero, ``{LaSR: A supple multi-connectivity scheduler for multi-RAT
  OFDMA systems},'' \emph{IEEE Transactions on Mobile Computing}, vol.~19,
  no.~3, pp. 624--639, 2018.

\bibitem{chen2017performance}
S.~Chen, T.~Zhao, H.-H. Chen, Z.~Lu, and W.~Meng, ``{Performance analysis of
  downlink coordinated multipoint joint transmission in ultra-dense
  networks},'' \emph{IEEE Network}, vol.~31, no.~5, pp. 106--114, 2017.

\bibitem{chetlur2018coverage}
V.~V. Chetlur and H.~S. Dhillon, ``{Coverage analysis of a vehicular network
  modeled as Cox process driven by Poisson line process},'' \emph{IEEE
  Transactions on Wireless Communications}, vol.~17, no.~7, pp. 4401--4416,
  2018.

\bibitem{andrews2011tractable}
J.~G. Andrews, F.~Baccelli, and R.~K. Ganti, ``A tractable approach to coverage
  and rate in cellular networks,'' \emph{IEEE Transactions on communications},
  vol.~59, no.~11, pp. 3122--3134, 2011.

\bibitem{qian2017non}
L.~P. Qian, Y.~Wu, H.~Zhou, and X.~Shen, ``{Non-orthogonal multiple access
  vehicular small cell networks: Architecture and solution},'' \emph{IEEE
  Network}, vol.~31, no.~4, pp. 15--21, 2017.

\bibitem{jiao2022spectral}
K.~Yu, H.~Zhou, Z.~Tang, X.~Shen, and F.~Hou, ``{Deep reinforcement
  learning-based RAN slicing for UL/DL decoupled cellular V2X},'' \emph{IEEE
  Transactions on Wireless Communications}, vol.~21, no.~5, pp. 3523--3535,
  2021.

\bibitem{xu2023federated}
Y.~Xu, B.~Qian, K.~Yu, T.~Ma, L.~Zhao, and H.~Zhou, ``{Federated Learning Over
  Fully-Decoupled RAN Architecture for Two-tier Computing Acceleration},''
  \emph{IEEE Journal on Selected Areas in Communications}, 2023.

\bibitem{kousaridas2019multi}
A.~Kousaridas, C.~Zhou, D.~Mart{\'\i}n-Sacrist{\'a}n, D.~Garcia-Roger, J.~F.
  Monserrat, and S.~Roger, ``{Multi-connectivity management for 5G V2X
  communication},'' in \emph{2019 IEEE 30th Annual International Symposium on
  Personal, Indoor and Mobile Radio Communications (PIMRC)}.\hskip 1em plus
  0.5em minus 0.4em\relax IEEE, 2019, pp. 1--7.

\bibitem{moltchanov2018upper}
D.~Moltchanov, A.~Ometov, S.~Andreev, and Y.~Koucheryavy, ``{Upper bound on
  capacity of 5G mmWave cellular with multi-connectivity capabilities},''
  \emph{Electronics Letters}, vol.~54, no.~11, pp. 724--726, 2018.

\bibitem{sylla2022multi}
T.~Sylla, L.~Mendiboure, S.~Maaloul, H.~Aniss, M.~A. Chalouf, and S.~Delbruel,
  ``{Multi-Connectivity for 5G Networks and Beyond: A Survey},''
  \emph{Sensors}, vol.~22, no.~19, p. 7591, 2022.

\bibitem{petrov2017dynamic}
V.~Petrov, D.~Solomitckii, A.~Samuylov, M.~A. Lema, M.~Gapeyenko,
  D.~Moltchanov, S.~Andreev, V.~Naumov, K.~Samouylov, M.~Dohler \emph{et~al.},
  ``{Dynamic multi-connectivity performance in ultra-dense urban mmWave
  deployments},'' \emph{IEEE Journal on Selected Areas in Communications},
  vol.~35, no.~9, pp. 2038--2055, 2017.

\bibitem{giordani2016multi}
M.~Giordani, M.~Mezzavilla, S.~Rangan, and M.~Zorzi, ``{Multi-connectivity in
  5G mmWave cellular networks},'' in \emph{2016 Mediterranean Ad Hoc Networking
  Workshop (Med-Hoc-Net)}.\hskip 1em plus 0.5em minus 0.4em\relax IEEE, 2016,
  pp. 1--7.

\bibitem{sattar2019spectral}
Z.~Sattar, J.~V. Evangelista, G.~Kaddoum, and N.~Batani, ``{Spectral efficiency
  analysis of the decoupled access for downlink and uplink in two-tier
  network},'' \emph{IEEE Transactions on Vehicular Technology}, vol.~68, no.~5,
  pp. 4871--4883, 2019.

\bibitem{shafie2020multi}
A.~Shafie, N.~Yang, and C.~Han, ``{Multi-connectivity for indoor terahertz
  communication with self and dynamic blockage},'' in \emph{ICC 2020-2020 IEEE
  International Conference on Communications (ICC)}.\hskip 1em plus 0.5em minus
  0.4em\relax IEEE, 2020, pp. 1--7.

\bibitem{kibria2018stochastic}
M.~G. Kibria, K.~Nguyen, G.~P. Villardi, W.-S. Liao, K.~Ishizu, and F.~Kojima,
  ``{A stochastic geometry analysis of multiconnectivity in heterogeneous
  wireless networks},'' \emph{IEEE Transactions on Vehicular Technology},
  vol.~67, no.~10, pp. 9734--9746, 2018.

\bibitem{kamble2009efficient}
V.~Kamble, S.~Kalyanasundaram, V.~Ramachandran, and R.~Agrawal, ``{Efficient
  resource allocation strategies for multicast/broadcast services in 3GPP long
  term evolution single frequency networks},'' in \emph{2009 IEEE Wireless
  Communications and Networking Conference}.\hskip 1em plus 0.5em minus
  0.4em\relax IEEE, 2009, pp. 1--6.

\bibitem{simsek2019multiconnectivity}
M.~Simsek, T.~H{\"o}{\ss}ler, E.~Jorswieck, H.~Klessig, and G.~Fettweis,
  ``{Multiconnectivity in multicellular, multiuser systems: A matching-based
  approach},'' \emph{Proceedings of the IEEE}, vol. 107, no.~2, pp. 394--413,
  2019.

\bibitem{chiu2013stochastic}
S.~N. Chiu, D.~Stoyan, W.~S. Kendall, and J.~Mecke, \emph{{Stochastic geometry
  and its applications}}.\hskip 1em plus 0.5em minus 0.4em\relax John Wiley \&
  Sons, 2013.

\bibitem{abdulqader2015performance}
A.~Abdulqader~Hussein, T.~A. Rahman, and C.~Y. Leow, ``{Performance evaluation
  of localization accuracy for a log-normal shadow fading wireless sensor
  network under physical barrier attacks},'' \emph{Sensors}, vol.~15, no.~12,
  pp. 30\,545--30\,570, 2015.

\bibitem{chetlur2019coverage}
{Chetlur, Vishnu Vardhan and Dhillon, Harpreet S}, ``{Coverage and rate
  analysis of downlink cellular vehicle-to-everything (C-V2X) communication},''
  \emph{IEEE Transactions on Wireless Communications}, vol.~19, no.~3, pp.
  1738--1753, 2019.

\bibitem{xu2016wireless}
C.~Xu, M.~Sheng, V.~S. Varma, T.~Q. Quek, and J.~Li, ``{Wireless service
  provider selection and bandwidth resource allocation in multi-tier HCNs},''
  \emph{IEEE Transactions on Communications}, vol.~64, no.~12, pp. 5108--5124,
  2016.

\bibitem{yang2015coverage}
X.~Yang and A.~O. Fapojuwo, ``{Coverage probability analysis of heterogeneous
  cellular networks in Rician/Rayleigh fading environments},'' \emph{IEEE
  Communications Letters}, vol.~19, no.~7, pp. 1197--1200, 2015.

\bibitem{jia2017downlink}
M.~Jia, Z.~Yin, Q.~Guo, G.~Liu, and X.~Gu, ``{Downlink design for spectrum
  efficient IoT network},'' \emph{IEEE Internet of Things Journal}, vol.~5,
  no.~5, pp. 3397--3404, 2017.

\bibitem{moltchanov2012distance}
D.~Moltchanov, ``{Distance distributions in random networks},'' \emph{Ad Hoc
  Networks}, vol.~10, no.~6, pp. 1146--1166, 2012.

\bibitem{berrar2018bayes}
D.~Berrar, ``Bayes’ theorem and naive bayes classifier,'' \emph{Encyclopedia
  of Bioinformatics and Computational Biology: ABC of Bioinformatics}, vol.
  403, p. 412, 2018.

\bibitem{stoyan2013stochastic}
D.~Stoyan, W.~S. Kendall, S.~N. Chiu, and J.~Mecke, \emph{{Stochastic geometry
  and its applications}}.\hskip 1em plus 0.5em minus 0.4em\relax John Wiley \&
  Sons, 2013.

\bibitem{3gpp.36.331}
``{Study on evaluation methodology of new vehicle-to-everything (V2X) use cases
  for LTE and NR},'' {Sophia Antipolis Cedex, France, 3GPP}, Tech. Rep. 37.885
  v15.2.0, Dec 2018.

\bibitem{3gpp.36.819}
``{Coordinated multi-point operation for LTE physical layer aspects},'' {Sophia
  Antipolis Cedex, France, 3GPP}, Tech. Rep. 36.819 v11.2.0, Sep 2013.

\bibitem{elshaer2016downlink}
H.~Elshaer, M.~N. Kulkarni, F.~Boccardi, J.~G. Andrews, and M.~Dohler,
  ``Downlink and uplink cell association with traditional macrocells and
  millimeter wave small cells,'' \emph{IEEE Transactions on Wireless
  Communications}, vol.~15, no.~9, pp. 6244--6258, 2016.

\end{thebibliography}

\begin{IEEEbiography}[{\includegraphics[width=1in,height=1.25in,clip,keepaspectratio]{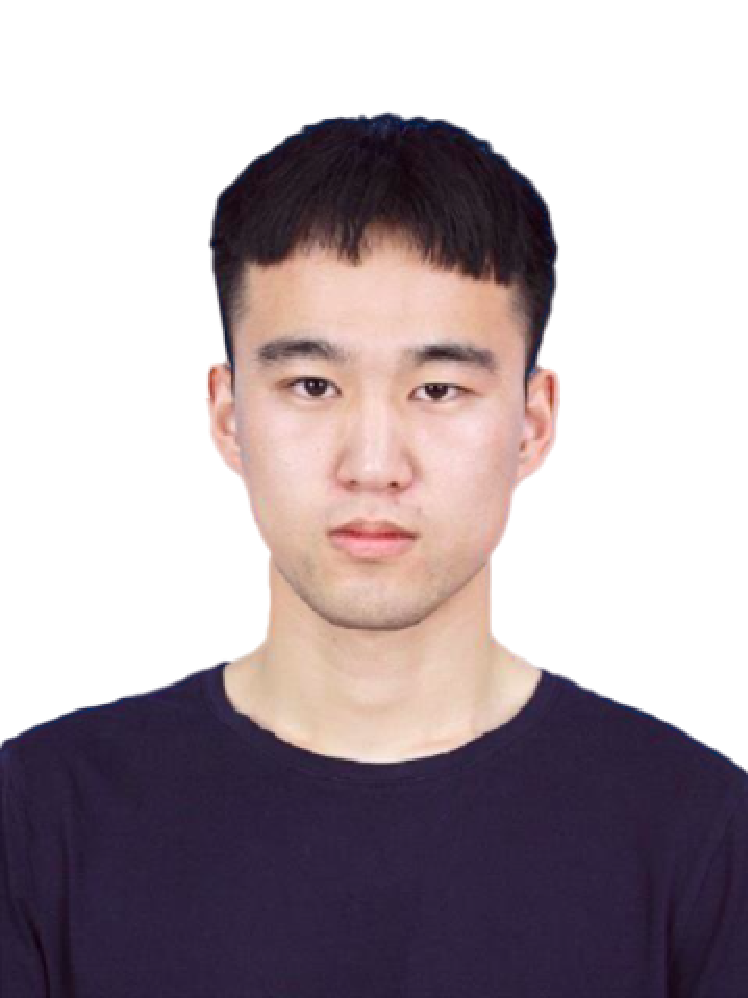}}]{Luofang Jiao} (Student Member, IEEE) received the B.S. degree in detection guidance and control technology from the University of Electronic Science and Technology of China, Chengdu, China, in 2020. He is currently working toward the Ph.D. degree with the School of Electronic Science and Engineering, Nanjing University, Nanjing, China. His research interests include uplink/downlink decoupled access, C-V2X, and heterogeneous networks.
\end{IEEEbiography}

\begin{IEEEbiography}[{\includegraphics[width=1in,height=1.25in,clip,keepaspectratio]{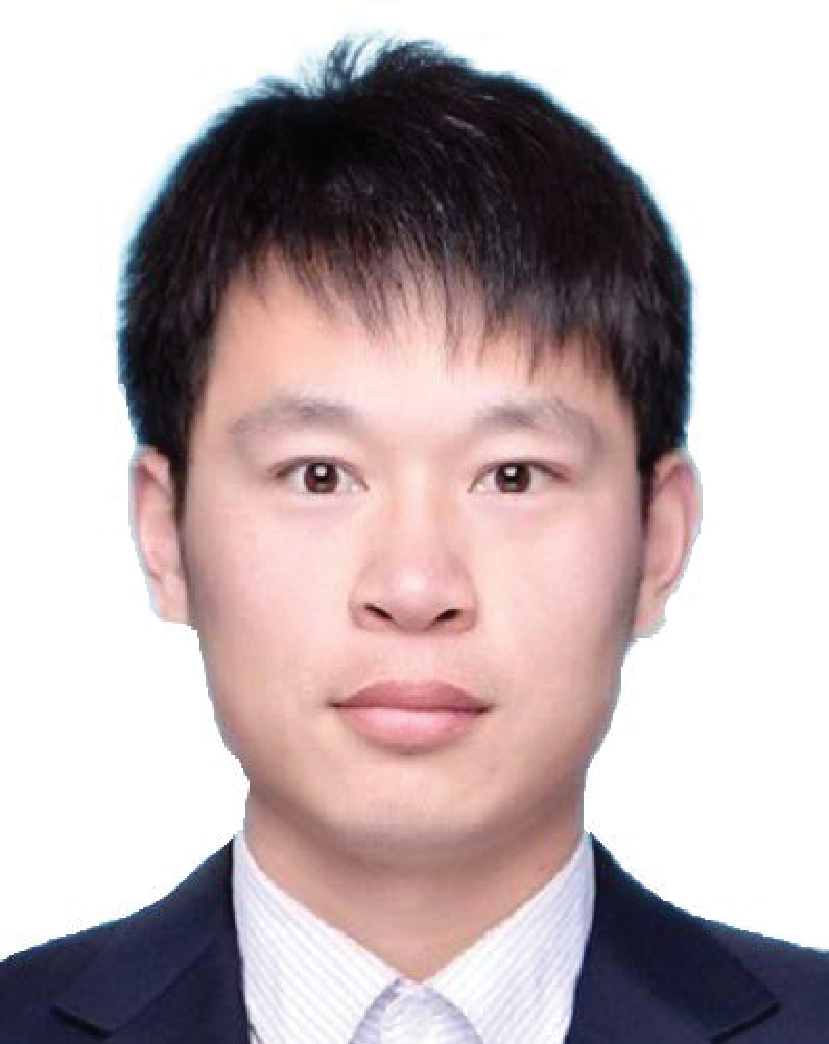}}]{Jiwei Zhao} (Student Member, IEEE) received the M.S. degree in information and communication system from Xidian University, Xi'an, China, in 2018. He is currently working toward the Ph.D. degree with the School of Electronic Science and Engineering, Nanjing University, Nanjing, China. He won the first prize in the 2016 CCF (China Computer Federation) China Big Data and Cloud Computing Intelligence Contest. His research interests include fully-decoupled RAN architecture, coordinated multi-point, and machine learning applications for wireless communication.
\end{IEEEbiography}

\begin{IEEEbiography}[{\includegraphics[width=1in,height=1.25in,clip,keepaspectratio]{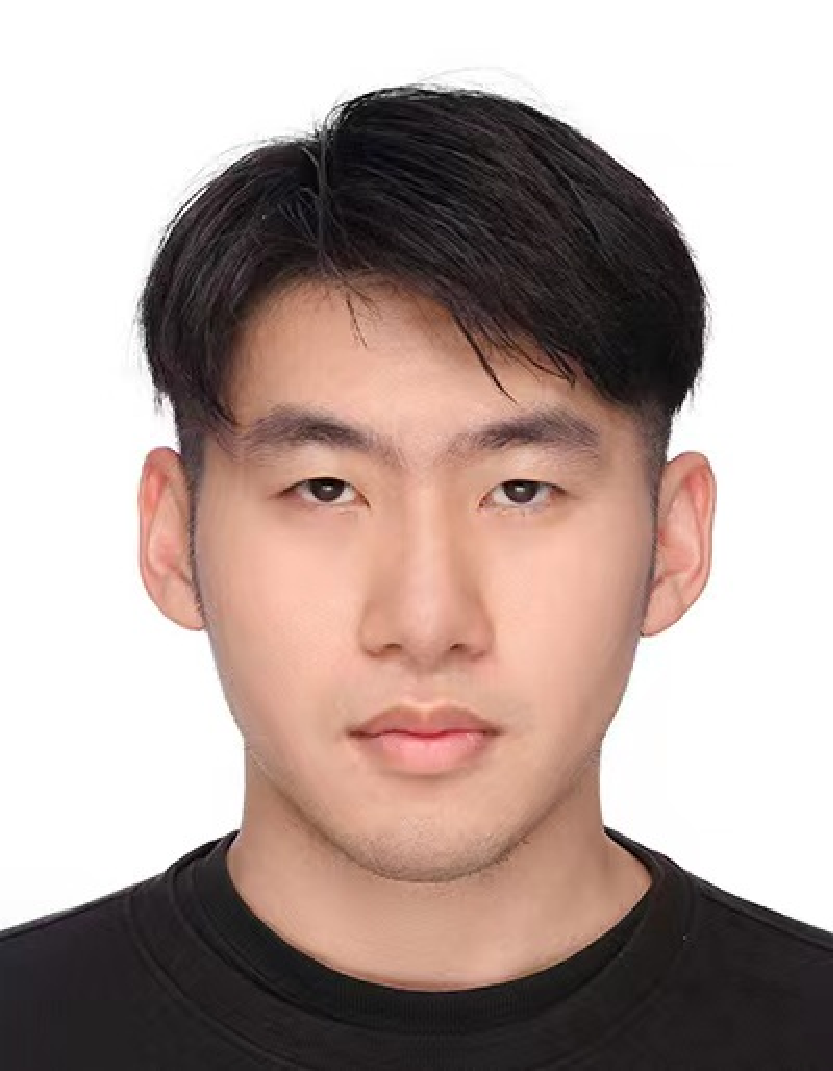}}]{Yunting Xu}
(Student Member, IEEE) received the B.S. degree in communication engineering from Nanjing University, Nanjing, China, in 2017, where he is currently pursuing the Ph.D. degree with the School of Electronic Science and Engineering. He mainly focuses on the dynamic resource management and networking optimization in the field of emerging wireless networks.
\end{IEEEbiography}

\begin{IEEEbiography}[{\includegraphics[width=1in,height=1.25in,clip,keepaspectratio]{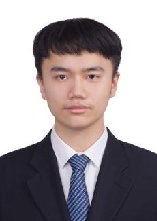}}]{Tianqi Zhang}
	(Student Member, IEEE) received the B.S. degree in electronic information science and technology from Nanjing University, Nanjing, China, in 2021, where he is currently pursuing the Ph.D. degree with the School of Electronic Science and Engineering. He mainly focuses on the FD-RAN, V2X, and machine learning in the field of emerging wireless networks.
\end{IEEEbiography}

\begin{IEEEbiography}[{\includegraphics[width=1in,height=1.25in,clip,keepaspectratio]{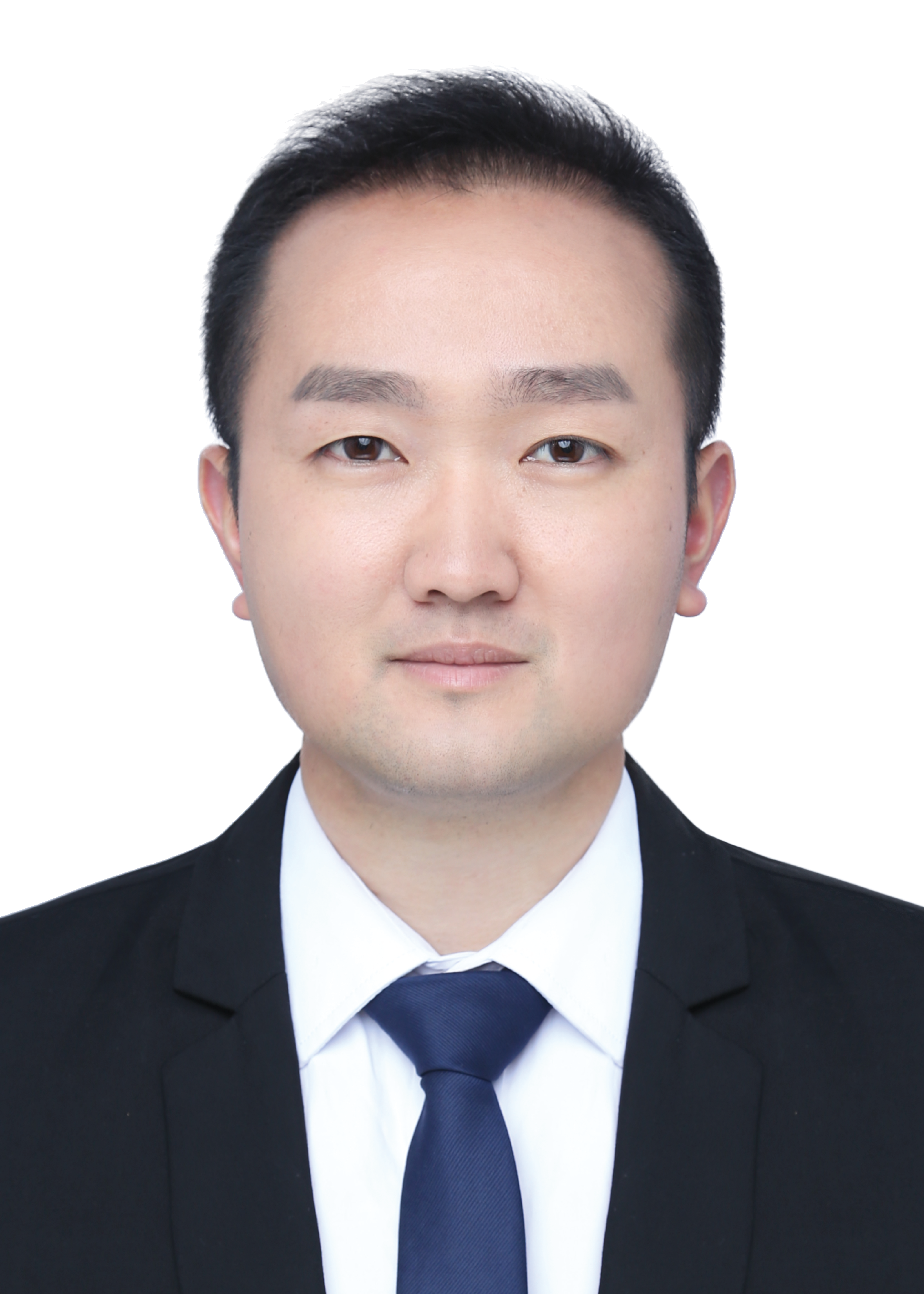}}]{Haibo Zhou} (Senior Member, IEEE) received the Ph.D. degree in information and communication engineering from Shanghai Jiao Tong University, Shanghai, China, in 2014. From 2014 to 2017, he was a Postdoctoral Fellow with the Broadband Communications Research Group, Department of Electrical and Computer Engineering, University of Waterloo. He is currently a Full Professor with the School of Electronic Science and Engineering, Nanjing University, Nanjing, China. He was elected as an IET fellow in 2022, highly cited researcher by Clarivate Analytics in 2022 \& 2020. He was a recipient of the 2019 IEEE ComSoc Asia–Pacific Outstanding Young Researcher Award, 2023-2024 IEEE ComSoc Distinguished Lecturer, and 2023-2025 IEEE VTS Distinguished Lecturer. He served as Track/Symposium CoChair for IEEE/CIC ICCC 2019, IEEE VTC-Fall 2020, IEEE VTC-Fall 2021, WCSP 2022, IEEE GLOBECOM 2022, IEEE ICC 2024. He is currently an Associate Editor of the IEEE Transactions on Wireless Communications, IEEE Internet of Things Journal, IEEE Network Magazine, and Journal of Communications and Information Networks. His research interests include resource management and protocol design in B5G/6G networks, vehicular ad hoc networks, and space-air-ground integrated networks.
\end{IEEEbiography}

\begin{IEEEbiography}[{\includegraphics[width=1in,height=1.25in,clip,keepaspectratio]{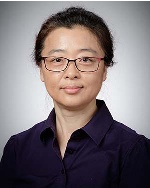}}]{Dongmei Zhao} (Senior Member, IEEE) received the B.S.degree in wireless communication from Northern Jiaotong University (currently, Beijing Jiaotong University), Beijing, China, in 1992, and the Ph.D degree from the Department of Electrical and Computer Engineering, University of Waterloo, Waterloo, ON, Canada, in June 2002. In July 2002, she joined the Department of Electrical and Computer Engineering, McMaster University, where she is a Full Professor. From April 2004 to March 2009, she was an Adjunct Assistant Professor with the Department of Electrical and Computer Engineering, University of Waterloo. Her current research areas are mainly in mobile computation offloading, energy efficient wireless networking, and vehicular communication networks. She is a Co-Chair of the Mobile and Wireless Networks Symposium of IEEE GLOBECOM Conference 2020, the Wireless Networking Symposium in IEEE GLOBECOM Conference 2007, the Green Computing, Networking, and Communications Symposium in International Conference on Computing, Networking and Communications 2020, the technical program committee for IEEE International Workshop on Computer Aided Modeling and Design of Communication Links and Networks 2016, the General Symposium of the International Wireless Communications and Mobile Computing (IWCMC) Conference 2007, and a co-chair of the Vehicular Networks Symposium of IWCMC from 2012 to 2019. He is an Associate Editor of the IEEE INTERNET OF THINGS JOURNAL. She served as an Associate Editor for the IEEE TRANSACTIONS ON VEHICULAR TECHNOLOGY from 2007 to 2017. She also served as an Editor for EURASIP Journal on Wireless Communications and Networking and Journal of Communications and Networks. She has been in Technical Program Committee of many international conferences in her fields. She is a Professional Engineer of Ontario.
\end{IEEEbiography}

\end{document}